\documentclass[journal]{IEEEtran}

\newcommand{\T}{{\top}}

\usepackage{amssymb}
\usepackage[english]{babel}
\usepackage[utf8x]{inputenc}
\usepackage{amsmath}
\usepackage[overload]{empheq}
\usepackage{framed}
\usepackage{marginnote}

\usepackage{caption}

\usepackage[numbers,sort&compress]{natbib}

\usepackage{lipsum}

\usepackage{mathtools}
\usepackage{cuted}

\allowdisplaybreaks

\usepackage{todonotes}
\usepackage{multirow,booktabs}
\usepackage{graphicx}
\usepackage{amsfonts}
\graphicspath{{images/}}
\usepackage{subfigure}
\usepackage[justification=centering]{caption}
\usepackage{epstopdf}
\usepackage{gensymb}
\usepackage[flushleft]{threeparttable}
\usepackage{balance} 
\usepackage{bm}

\usepackage{float}

\usepackage{algorithmic}
\usepackage{algorithm}

\allowdisplaybreaks

\newtheorem{remark}{Remark}

\makeatletter
\def\RemoveSpaces#1{\zap@space#1 \@empty}
\makeatother

\begin{document}


%
%

\title{Nonlinear Double-Capacitor Model for Rechargeable Batteries: Modeling, Identification and Validation}

\author{
	\vskip 1em
	Ning Tian, \emph{Student Member}, \emph{IEEE},
	Huazhen Fang, \emph{Member}, \emph{IEEE}, Jian Chen, \emph{Senior Member}, \emph{IEEE}, and Yebin Wang, \emph{Senior Member}, \emph{IEEE}

	\thanks{This work was supported in part by the National Science Foundation  under Awards CMMI-1763093 and CMMI-1847651.
		
		Ning Tian and Huazhen Fang are with the Department of Mechanical Engineering, University of Kansas, Lawrence, KS 66045, USA (e-mail: ning.tian@ku.edu, fang@ku.edu).

Jian Chen is with the State Key Laboratory of Industrial Control Technology, College of Control Science and Engineering, Zhejiang University, Hangzhou 310027, China, and also with the Ningbo Research Institute, Zhejiang University, Ningbo 315100, China (e-mail: jchen@zju.edu.cn).

Yebin Wang is with the Mitsubishi Electric Research Laboratories, Cambridge,
MA 02139, USA (e-mail:l yebinwang@ieee.org).
		
	}
}


\maketitle

\begin{abstract}
This paper proposes a new equivalent circuit model for rechargeable batteries by modifying a double-capacitor model proposed in the literature.  It is known that the original model can  address the rate capacity effect  and energy recovery effect inherent to  batteries better than other models. However, it is a purely linear model and includes no representation of a battery's nonlinear phenomena. Hence, this work transforms the original model by introducing a nonlinear-mapping-based voltage source and a serial RC circuit.  The modification is justified by an analogy with the single-particle model. Two offline parameter estimation approaches, termed 1.0 and 2.0, are designed for the   new  model to deal with the scenarios of constant-current and variable-current charging/discharging, respectively. In particular, the 2.0 approach proposes the notion of Wiener system identification based on the maximum {\em a posteriori} estimation, which allows all the parameters to be estimated in one shot while overcoming the nonconvexity or local minima issue to obtain physically reasonable estimates. 
An extensive experimental evaluation shows that the proposed model offers excellent accuracy and predictive capability. A comparison against the Rint   and Thevenin models further points to its superiority. With high fidelity and low mathematical complexity, this model is beneficial for various real-time battery management applications. 
\end{abstract} 



\definecolor{limegreen}{rgb}{0.2, 0.8, 0.2}
\definecolor{forestgreen}{rgb}{0.13, 0.55, 0.13}
\definecolor{greenhtml}{rgb}{0.0, 0.5, 0.0}

\section{Introduction}\label{Sec:Introduction}

\IEEEPARstart{R}{echargeable} batteries have seen an ever-increasing use in today's industry and society  as power sources for systems of different scales, ranging from consumer electronic devices to electric vehicles and smart grid.  This trend has motivated a growing body of research  on advanced battery management algorithms, which are aimed to ensure the performance, safety and life of  battery systems. Such algorithms generally require mathematical models that can well characterize  a battery's dynamics. This has stimulated significant attention in battery modeling   during the past years, with the current literature offering a plethora of results. 

There are two main types of battery models:  1) electrochemical models that build on electrochemical principles to describe the electrochemical reactions and physical phenomena inside a battery during charging/discharging, and  2) equivalent circuit models (ECMs) that  replicate a battery's current-voltage characteristics using electrical circuits  made  of  resistors, capacitors and voltage sources.  With structural simplicity, the latter ones provide great computational efficiency, thus  more suitable    for real-time battery management. However, as the other side of the coin,  the simple circuit-based structures also imply a difficulty to capture a battery's dynamic behavior at a high accuracy. Therefore, this article aims to develop a new ECM that offers not only structural parsimony but also high fidelity, through transforming an existing model in~\cite{Johnson:NREL:2000}. The work will systematically investigate the model construction, parameter identification, and experimental validation.

\subsection{Literature Review}

\subsubsection{Review of Battery Modeling} As mentioned above, the electrochemical models and ECMs constitute the majority of the battery models available today.
The electrochemical modeling approach seeks to characterize the  physical and chemical mechanisms  underlying the charging/discharging processes.
One of the best-known electrochemical models is  the Doyle-Fuller-Newman model, which  describes the concentrations and transport of lithium ions together with the distribution of separate  potential in porous electrodes and electrolyte~\cite{DFN:JTES:1993,Forman:JPS:2012,Chaturvedi:CSM:2010}. While delineating and reproducing a battery's behavior accurately, 
this model, like many others of similar kind, involves many partial differential equations and causes  high computational costs. This has driven the development of some simplified versions, e.g., the single-particle model (SPM)~\cite{Chaturvedi:CSM:2010,Guo:JES:2011}, and various model reduction methods, e.g.,~\cite{Northrop:JES:2014,Zou:TCST:2016,Hu:Mechatronics:2018}, toward more efficient computation.



By contrast, the ECMs  are generally considered as  more competitive   for real-time battery monitoring and control, having found their way into various battery management systems.  The first ECM to our knowledge is the Randles model proposed in the 1940s~\cite{Randles:DFS:1947}. It reveals a lead-acid battery's ohmic and reactive (capacitive and inductive) resistance, demonstrated in the electrochemical reactions and contributing to various phenomena of voltage dynamics, e.g., voltage drop, recovery and associated transients. This model has become  a de facto standard for interpreting battery data obtained from  electrochemical impedance spectroscopy (EIS)~\cite{Zia:Springer:2016}. It also provides a basis for building diverse ECMs to grasp a battery's voltage dynamics during charging/discharging. 
Adding a voltage source representing the open-circuit voltage (OCV) to the Randles model, one can obtain the popular Thevenin model~\cite{Mousavi:RSER:2014,He:Energies:2011,Plett:Artech:2015}. 
The Thevenin model without the resistance-capacitance (RC) circuit is called as the Rint model, which includes an ideal voltage source with a series resistor~\cite{He:Energies:2011}. If   more than one RC circuit is added to the Thevenin model, it becomes the dual polarization (DP) model that is  capable of capturing multi-time-scale  voltage transients during charging/discharging~\cite{He:Energies:2011}.

The literature  has also reported a few modifications  of the Thevenin model to better characterize a battery's dynamics. Generally, they are based on two approaches. The first one aims to describe a battery's voltage more accurately by incorporating certain phenomena, e.g., hysteresis, into the voltage dynamics, or through different parameterizations of   OCV with respect to the state of charge (SoC)~\cite{Weng:JPS:2014,Lin:JPS:2014,gholizadeh2014estimation,Perez:TVT:2017,Fang:CEP:2014,Plett:JPS:2004,Hu:JPS:2012}. 
Some literature also models the resistors and capacitors as dependent on   SoC, as well as some other factors like the temperature or rate and direction of the current loads in order to   improve the accuracy of battery voltage prediction~\cite{li2018practical,lee2018temperature}. 
The second approach sets the focus on improving the runtime prediction for batteries.  In~\cite{Chen:TEC:2006}, a battery's capacity change due to cycle and temperature  is considered and parameterized, and the dependence of resistors and capacitors on SoC also characterized. A similar investigation is made in~\cite{Kim:TEC:2011} to improve the Thevenin model, which proposes to capture the nonlinear change of a battery's capacity with respect to the current loads. 

An ECM that shows emerging importance is a  double-capacitor model~\cite{Johnson:NREL:2000,Johnson:JPS:2002}. It consists of two capacitors configured in parallel, which correspond to an electrode's bulk inner part and surface region, respectively, and  can describe the process of charge diffusion and storage in a battery's electrode~\cite{Fang:TCST:2017}. 
 Compared to  the Thevenin model, this   circuit structure allows  the rate capacity effect and charge recovery effect to be captured, making the model an attractive choice for charging control~\cite{Fang:TCST:2017,Fang:JES:2018}. However, based on a  purely linear circuit, this model is unable to grasp   nonlinear phenomena innate to a battery---for instance, the nonlinear SoC-OCV relation is beyond its descriptive capability---and thus has its applicability limited. The presented work is   motivated to remove this limitation by revamping the model's structure. The effort will eventually lead to a new  ECM that, for the first time, can capture the charge diffusion within a battery's electrode and its nonlinear voltage behavior simultaneously.
 

\subsubsection{Review of Battery Model Identification}

A key problem associated with battery modeling is parameter identification, which pertains to extracting the unknown model parameters from the measurement data. 
Due to its importance, recent years have seen  a growth of research. The existing methods can be divided into two main categories, experiment-based and data-based. The first category conducts experiments of charging, discharging or EIS and utilizes the experimental data to read a model's parameters. It is pointed out in~\cite{schweighofer2003modeling,abu2004rapid} that the transient voltage responses under constant- or pulse-current charging/discharging can be leveraged to estimate the  resistance, capacitance and time constant parameters of the   Thevenin model. In addition, the relation  between   SoC and OCV is a defining characteristic of a battery's dynamics. It can be experimentally identified by charging or discharging a battery using a very small current~\cite{dubarry2007development}, or alternatively, using a current of normal magnitude but intermittently (with a sufficiently long rest period applied between two discharging operations)~\cite{he2011state,tian2014modified}. The EIS experiments have also been widely used to identify a battery's   impedance properties~\cite{mauracher1997dynamic,goebel2008prognostics,Birkl:HEVC:2013}. While    involving  basic data analysis, the methods of this category generally put  emphasis on the design of experiments. In a departure, the second category goes deeper into understanding the model-data relationship and pursues   data-driven parameter estimation. It can   enable provably correct identification even for complex models, thus often acknowledged as   better at extracting the potential of data. It is proposed in~\cite{Hu:JPS:2013} to identify the Thevenin model by solving a set of linear and polynomial equations. Another popular means is to formulate   model-data fitting problems and solve them using least squares or other optimization methods to estimate the parameters~\cite{he2016parameter,yang2016improved,tian2017parameter,feng2015online,prasad2013model,sitterly2011enhanced}. When  considering  more complicated electrochemical models, the identification usually involves large-size nonlinear nonconvex optimization problems. In this case, particle swarm optimization and genetic algorithms are often exploited to search for the best parameter estimates~\cite{Forman:JPS:2012,zhang2014multi,rahman2016electrochemical,yu2017model}. A recent study presents an adaptive-observer-based parameter estimation scheme for an electrochemical model~\cite{Limoge:TCST:2018}. While the above works focus on identification of physics-based models, data-driven black-box identification is also  examined in~\cite{Hu:JPS:2011,Li:IJER:2014,Relan:TCST:2017}, which construct linear state-space models {\em via } subspace identification or nonparametric frequency domain analysis. A topic related with identification is experiment design, which is to find out the best input sequences to excite a battery to maximize the parameter identifiability. In~\cite{Rothenberger:JES:2015,Park:ACC:2018}, optimal input design is performed by maximizing the Fisher information matrices---an identifiability metric---involved in the identification of the Thevenin model and the SPM, respectively.

The presented work is also related with the literature on Wiener system identification, because the model to be developed  has a Wiener-type structure featuring a linear dynamic subsystem in cascade with a static nonlinear  subsystem.  Wiener systems are an important    subject in the field of parameter identification, and a reader is referred to~\cite{giri2010block} for a collection of recent studies.  Wiener system identification    based on   maximum likelihood (ML) estimation is investigated in~\cite{Hagenblad:AUTO:2008,Vanbeylen:TSP:2009}, which shows significant promises. However, the optimization procedure resulting from the ML formulation can easily converge to local minima    due to  the presence of the nonlinear subsystem.  This  hence yields  a motivation to   enhance the notion of ML-based   identification in this work to achieve more effective battery parameter estimation.


\subsection{Statement of Contributions} 


This work presents the following contributions.

\begin{itemize}

\item  A new   ECM, named the {\em  nonlinear double-capacitor (NDC) model}, is developed. By design, it transforms the linear double-capacitor model in~\cite{Johnson:NREL:2000} by   coupling it with a nonlinear  circuit mimicking a battery's voltage behavior. With this pivotal change, the NDC model introduces two   advantages over existing ECMs. First, it can   simulate not only  the charge diffusion characteristic of a battery's electrochemical dynamics,  but also the critical nonlinear  electrical phenomena. This unique feature guarantees the model's better   accuracy, which comes at only a  very slight increase in model complexity. Second, the NDC model can be interpreted  as a circuit-based approximation of the SPM. This further  justifies its soundness while inspiring a refreshed look at the connections between the   SPM and ECMs.


\item Parameter identification  is investigated for the proposed  model. This begins with a study  of the  constant-current charging/discharging scenario, with an identification approach, termed 1.0, developed by   fitting   parameters with the measurement data. Then, shifting the focus to the    scenario of variable-current charging/discharging, the study introduces a Wiener   perspective into the identification of the NDC model  due to its Wiener-type structure. A Wiener   identification approach is proposed for the NDC model based on maximum {\em a posteriori} (MAP) estimation, which is termed 2.0. Compared to the ML-based   counterparts in the literature, this new approach incorporates into the estimation   a prior distribution of the unknown parameters, which represents additional information  or prior knowledge and can help drive the parameter search toward  physically reasonable values. 



\item Experimental validation is performed to assess the proposed results. This involves  multiple experiments  about battery discharging under different kinds of current profiles and a comparison of the NDC model with the Rint and Thevenin models. The validation shows  the considerable accuracy and predictive capability of the NDC model, as well as the   effectiveness of  the 1.0 and 2.0  identification approaches.



\end{itemize}

\subsection{Organization}

The remainder of the paper is organized as follows. Section~\ref{Sec:Modeling} presents the construction of the   NDC model. Section~\ref{Sec:Parameter-Identification-1}   studies    parameter identification for the NDC model in the constant-current charging/discharging scenario. Inspired by Wiener system identification, Section~\ref{Sec:Parameter-Identification-2} proceeds to   develop  an MAP-based parameter estimation approach to identify the NDC model. Section~\ref{Sec:Experimental-Validation}  offers the experimental validation. Finally, Section~\ref{Sec:Conclusion} gathers concluding remarks and suggestions for future research.

\begin{figure}[t]
\captionsetup{justification = raggedright, singlelinecheck = false}
\centering
\vspace{-5mm}
\subfigure[]
{\includegraphics[trim = {72mm 70mm 122mm 50mm}, clip, width=0.38\textwidth]{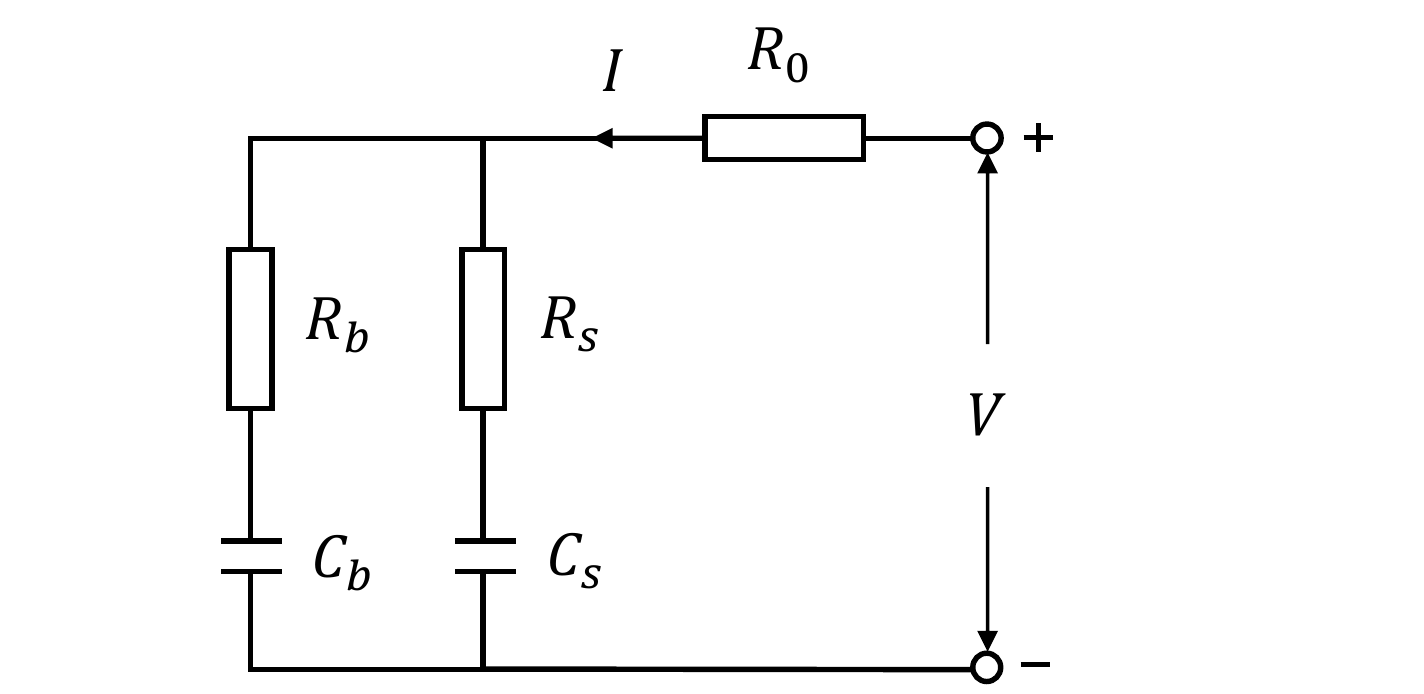}\label{Fig:traditional-RCM-diagram}}
\subfigure[]
{\includegraphics[trim = {80mm 70mm 80mm 38mm}, clip, width=0.47\textwidth]{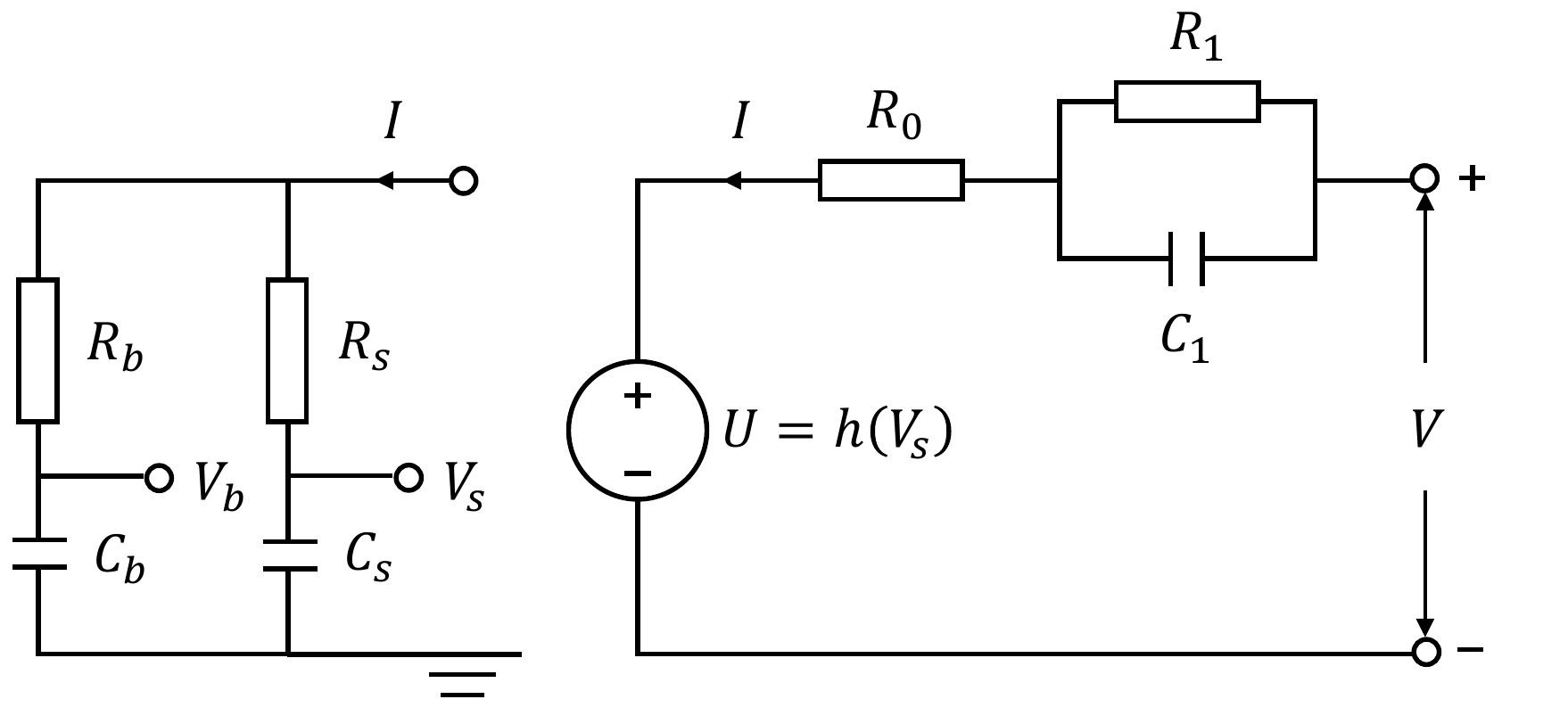}\label{Fig:modified-RCM-diagram}}
\vspace{-3mm}
\caption{(a) The original double-capacitor model; (b) the proposed NDC model.} 
\label{Fig:RCM-diagram}
\vspace{0mm}
\end{figure}

\section{{NDC} Model Development}\label{Sec:Modeling}

This section develops the   NDC model  and presents the mathematical equations governing its dynamic behavior. 

To begin with, let us review the original linear double-capacitor model proposed in~\cite{Johnson:NREL:2000}.
As shown in Figure~\ref{Fig:traditional-RCM-diagram}, this model includes two capacitors in parallel, $C_b$ and $C_s$, each   connected with a serial resistor, $R_b$ and $R_s$, respectively. The double-capacitor structure simulates a battery's electrode,  providing storage for electric charge, and the parallel connection between them allows  the  transport of charge within the electrode to be described. Specifically, one can consider the $R_s$-$C_s$ circuit  as corresponding to  the electrode surface region exposed to the electrolyte; the $R_b$-$C_b$ circuit represents an analogy of the bulk inner part of the electrode. 
As such, this model has the following features: 
\begin{itemize}
\item $C_b \gg C_s$ and $R_b \gg R_s$;

\item $C_b$ is where the majority of the charge is stored, and $R_b$-$C_b$ accounts for   low-frequency responses during charging/discharging;

\item  $C_s$ is much smaller, and its voltage changes at much faster rates than that of $C_b$ during charging/discharging, making   $R_s$-$C_s$    responsible for   high-frequency responses.
\end{itemize}
In addition, $R_0$ is included  to embody the electrolyte resistance.  This model was designed in~\cite{Johnson:NREL:2000} for high-power lithium-ion batteries, and its application can naturally extend to double-layer capacitors that are widely used in hybrid energy storage systems, e.g.,~\cite{Dey:TCST:2019}.

As pointed out in~\cite{Fang:TCST:2017}, the linear double-capacitor model can grasp the rate capacity effect, i.e., the total charge absorbed (or released) by a battery goes down with the increase in charging (or discharging) current. To see this, just notice that the terminal voltage $V$   mainly depends on  $V_s$ (the voltage across $C_s$), which changes faster than $V_b$ (the voltage across $C_b$). Thus, when the  current $I$ is large, the fast rise (or decline) of $V_s$ will make $V$ hit the cut-off threshold earlier than when $C_b$ has yet to be fully charged (or discharged). Another phenomenon that can be seized is the capacity and voltage  recovery effect. That is, the usable capacity and terminal voltage would increase upon the termination of discharging due to the migration of charge from $C_b$ to $C_s$.
However, this model by nature is a linear system, unable to describe a defining characteristic of batteries---the nonlinear dependence of   OCV on the SoC. It hence is effective only when a battery is restricted to operate  conservatively   within  some truncated SoC range that permits   a linear approximation of  the SoC-OCV curve.  

To overcome the above issue, the NDC model is  proposed, which is shown in Figure~\ref{Fig:modified-RCM-diagram}. It includes two  changes.  The primary one is to introduce  a voltage source $U$, which is   a nonlinear mapping of $V_s$, i.e.,  $U=h(V_s)$. Second, an RC circuit, $R_1$-$C_1$, is added in series to $U$. 
Next, let us justify the above modifications from a perspective of the SPM, a simplified electrochemical model that has recently attracted wide interest. 

\begin{figure}[t]
\captionsetup{justification = raggedright, singlelinecheck = false}
\centering
\includegraphics[trim = {50mm 10mm 70mm 20mm}, clip, width=0.48\textwidth]{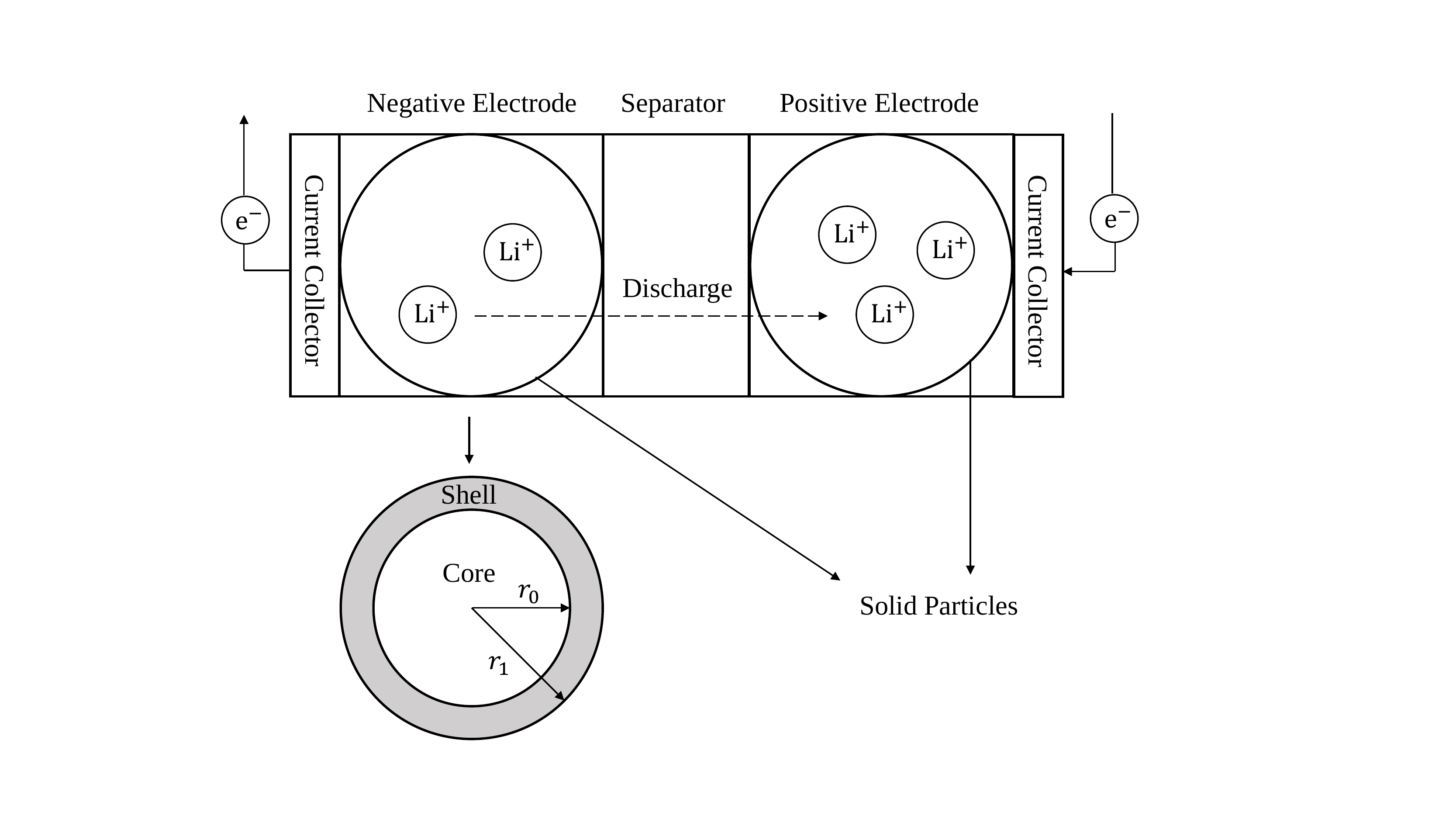}
\caption{The single-particle model (top), and a particle (bottom) subdivided into two volumes, core and shell, which correspond to $R_b$-$C_b$ and $R_s$-$C_s$, respectively.}
\label{Fig:Single-Particle-Model}
\end{figure}

Figure~\ref{Fig:Single-Particle-Model} gives a schematic diagram of the SPM.  
The SPM represents an electrode as a single spherical particle. It describes  the mass balance and diffusion of lithium ions in a particle during charging/discharging by Fick’s second law of diffusion in a spherical coordinate system~\cite{Guo:JES:2011}.  If subdividing a spherical particle into  two finite volumes, the bulk inner domain (core) and the near-surface domain (shell), one can simplify the diffusion of lithium ions between them as the charge transport between the capacitors of the double-capacitor model, as proven in~\cite{Fang:TCST:2017}. For SPM, the terminal voltage consists of three elements: the difference in the open-circuit potential of the positive and negative electrodes, the difference in the reaction overpotential, and the voltage across the film resistance~\cite{Chaturvedi:CSM:2010}. The open-circuit potential depends on the lithium-ion concentration in the surface region of the sphere, which is akin to the role of $V_s$ here. Therefore, it is appropriate as well as necessary to introduce a nonlinear function of $V_s$, i.e., $h(V_s)$, as an analogy to the open-circuit potential.
With $U=h(V_s)$, the NDC model can correctly show the influence of the charge state on the terminal voltage, while inheriting all the capabilities of the original model.

Furthermore, the NDC model also contains  an RC circuit, $R_1$-$C_1$, which, together with $R_0$, simulates the impedance-based part of the  voltage dynamics. Here, $R_0$ characterizes the linear kinetic aspect of the impedance, which relates to the ohmic resistance and solid electrolyte interface (SEI) resistance~\cite{mamun2018collective}; $R_1$-$C_1$ accounts for the  voltage transients  related with the charge transfer on the electrode/electrolyte interface and the ion mass diffusion in the battery~\cite{andre2011characterization}. This work finds that one RC circuit  can offer sufficient  fidelity, though it is possible to connect more RC circuits serially with $R_1$-$C_1$ to gain better accuracy. 


The dynamics of the NDC model can be expressed in the state-space form as follows:
\begin{subequations}\label{state-space-equation}
\begin{align}[left = \empheqlbrace\,] \label{state-equation}
\begin{bmatrix}
\dot{V}_{b}(t) \\ \dot{V}_s(t) \\ \dot{V}_1(t)
\end{bmatrix}
& =
A
\begin{bmatrix}
V_b(t) \\ V_s(t) \\V_1(t)
\end{bmatrix}
+
B
I(t), \\ \label{measurement-equation}
V(t) & = h(V_s(t))-V_1(t)+R_0I(t),
\end{align}
\end{subequations}
where
\begin{align*}
A = \begin{bmatrix}
\frac{-1}{C_b(R_b+R_s)} & \frac{1}{C_b(R_b+R_s)} & 0\\
\frac{1}{C_s(R_b+R_s)} & \frac{-1}{C_s(R_b+R_s)} & 0\\
0 & 0& \frac{-1}{R_1C_1}
\end{bmatrix}, \ 
B = \begin{bmatrix}
\frac{R_s}{C_b(R_b+R_s)}\\ \frac{R_b}{C_s(R_b+R_s)} \\ \frac{-1}{C_1}
\end{bmatrix}.
\end{align*}
In above, $I>0$  for charging, $I<0$ for discharging, and $V_1$ refers to the voltage across the $R_1$-$C_1$ circuit. One can parameterize 
$h(V_s)$ as a polynomial. A fifth-order polynomial is empirically selected here:
\begin{align}\label{hVs-original-function} \nonumber
h(V_s) = \alpha_0+\alpha_1V_s +\alpha_2V_s^2 +\alpha_3V_s^3 +\alpha_4V_s^4 +\alpha_5V_s^5,
\end{align}
where $\alpha_i$ for $i=0,1,\ldots,5$ are coefficients. Note that $h(V_s)$ should be lower and upper bounded, depending on a battery's operating voltage range. This implies that $V_b$ and $V_s$ must also be bounded. For any bounds selected for them, it is always possible to find out a set of coefficients $\alpha_i$'s to satisfy $h(\cdot)$. Hence,  one can straightforwardly normalize $V_b$ and $V_s$ to let them lie between 0 V and 1 V, without loss of generality. In other words, $V_b=V_s=1$ V at full charge (${\rm SoC}=1$)  and that $V_b=V_s=0$ V for full depletion (${\rm SoC}=0$). Following this setting,   SoC is given by
\begin{align}\label{SoC-equation} 
{\rm SoC}=\frac{Q_a}{Q_t}=\frac{C_b V_b+C_s V_s}
{C_b +C_s },
\end{align}
where  $Q_t = C_b +C_s$  denotes the total capacity, and $Q_a = C_b V_b+C_s V_s$  the available capacity, respectively. It is easy to verify that the SoC's dynamics is governed by
\begin{align} \label{SoC-Dynamics}
\dot{\rm SoC}=
\begin{bmatrix}
\frac{C_b}{C_b+C_s}&\frac{C_s}{C_b+C_s}&0
\end{bmatrix}
\begin{bmatrix}
\dot{V}_{b}\\ \dot{V}_s \\ \dot{V}_1
\end{bmatrix}
=\frac{1}{Q_t}I.
\end{align} 
Meanwhile, it is worth noting that  the SoC-OCV function would share  the same form with $h(\cdot)$. To see this point, recall that   OCV refers to the terminal voltage when the battery is at equilibrium without current load. For the NDC model, the equilibrium happens when $V_b=V_s$, $V_1=0$ V and $I=0$ A, and in this case, $V_s = \mathrm{SoC}$ according to~\eqref{SoC-equation}, and $\mathrm{OCV} = h(V_s)$. This suggests that ${\rm OCV} = h({\rm SoC})$. In addition, the internal resistance $R_0$ is also assumed to be SoC-dependent following the recommendation in~\cite{dubarry2009single}, taking  the form of
\begin{align}\label{R0-SoC}
R_0=\gamma_1+\gamma_2e^{-\gamma_3{\rm SoC}}+\gamma_4e^{-\gamma_5{\rm (1-SoC)}}.
\end{align}
The rest of this paper will center on developing parameter identification approaches to determine the model parameters using measurement data and apply  identified models to  experimental datasets to evaluate their predictive accuracy.

\section{Parameter Identification 1.0:  Constant-Current Charging/Discharging}\label{Sec:Parameter-Identification-1}
This section studies parameter identification for the NDC model when a constant current is applied to a battery. The  discharging case  is considered here  without loss of generality. In a two-step procedure, the $h(\cdot)$ function is identified first, and   the impedance and capacitance parameters estimated next. 



\subsection{Identification of $h(\cdot)$ }\label{Sec:SoC-OCV-identification}
The SoC-OCV relation of the NDC model is given by ${\rm OCV} = h({\rm SoC})$, as aforementioned in Section~\ref{Sec:Modeling}. Hence, one can identify $h(\cdot)$  by fitting it with a battery's SoC-OCV data. To obtain the SoC-OCV curve, one can discharge a battery using a small current (e.g.,  1/25 C-rate as suggested in~\cite{dubarry2007development}) from full to empty. 
In  this process, the terminal voltage $V$  can be taken as   OCV. Immediately one can see that $\alpha_0=\underline{V}$ and $\sum^5_{i=0}\alpha_i=\overline{V}$, where $\underline{V}$ and $\overline{V}$ are the minimum and maximum value of $V$ in the process. Therefore, ${\rm OCV}=h({\rm SoC})$ can be written as a function of $\alpha_i$ for $i=1,2,\ldots,4$ as follows:
\begin{align}\nonumber
{\rm OCV} = \underline{V}+\sum^{4}_{i=1}\alpha_i{\rm SoC}^i+
\left(\overline{V}-\underline{V}-\sum^4_{i=1}\alpha_i\right){\rm SoC}^5,
\end{align}
where $\rm OCV$ can be read directly from the terminal voltage measurements. By~\eqref{SoC-Dynamics},   SoC can be calculated using the coulomb counting method as follows:
\begin{align*}
{\rm SoC} = 1+\frac{1}{Q_t}It.
\end{align*}
From above, one can observe that $\alpha_i$ for $i=1,2,\ldots,4$ can be identified by solving a data fitting problem, which can be addressed as a linear least squares problem. The identification results are unique and can be easily obtained. Then with $\alpha_0=\underline{V}$ and $\alpha_5 =\overline{V}-\underline{V}- \sum^4_{i=1}\alpha_i$, the function $h(\cdot)$ becomes explicit and ready for use.

\subsection{Identification of Impedance and Capacitance}\label{Sec:Impedance-Identification}
Now, consider discharging the battery by a   constant current of normal magnitude to determine the impedance and capacitance parameters. The identification can be attained by expressing the terminal voltage  in terms of the parameters and then fitting it to the  measurement data. 
\subsubsection{Terminal Voltage Response Analysis}\label{Sec:Reparameterization}
Consider a battery left idling for a long period of time, and then discharge it using a constant current.  According to~\eqref{state-equation}, $V_s$ can be derived as
\begin{align}\label{original-Vs} \nonumber
V_s(t)&= V_s(0)+\frac{It}{C_b+C_s}+\frac{C_b(R_bC_b-R_sC_s)I}{(C_b+C_s)^2}\\
&\quad\quad \cdot\left[1-\exp\left({-\frac{C_b+C_s}{C_bC_s(R_b+R_s)}t}\right)\right],
\end{align}
where $V_s(0)$ is known to us as it can be accessed from $\rm{SoC}(0)$ when the battery is initially relaxed.  However, it is impossible to identify $C_b$, $R_b$, $C_s$ and $R_s$ altogether. This issue  can   be seen from~\eqref{original-Vs}, where $V_s$ depends on three   parameters, i.e., $1/ (C_b+C_s)$, ${C_b(R_bC_b-R_sC_s)}/{(C_b+C_s)^2}$ and $(C_b+C_s)/\left[{C_bC_s(R_b+R_s)}\right]$. Even if the three parameters are known, it is still not possible to extract all the four individual impedance and capacitance parameters from them due to the parameter redundancy. Therefore, one can sensibly assume  $R_s=0$, as recommended  in~\cite{sitterly2011enhanced}. This is a tenable assumption for the NDC model since  $R_s\ll R_b$ as aforementioned. As a result,~\eqref{original-Vs} reduces to
\begin{align}\label{final-Vs}
V_s(t) = V_s(0)+\beta_1It+\beta_2I\left(1-e^{-\beta_3t}  \right),
\end{align}
where 
\begin{align*}
\beta_1 = \frac{1}{C_b+C_s}, \ \
\beta_2  = \frac{R_bC_b^2}{(C_b+C_s)^2},\ \  
\beta_3  = \frac{C_b+C_s}{C_bC_sR_b}.
\end{align*}
Here, $\beta_1$ is known because $ Q_t$ has been calibrated by coulomb counting in  Section~\ref{Sec:SoC-OCV-identification}. When $\beta_2$ and $\beta_3$ are also available, $C_b$, $C_s$ and $R_b$ can be reconstructed as follows:
\begin{align*} 
C_b = \frac{\beta_2\beta_3}{\beta_1(\beta_1+\beta_2\beta_3)},\ 
C_s = \frac{1}{\beta_1+\beta_2\beta_3},\ 
R_b = \frac{1}{\beta_1\beta_3C_bC_s}.
\end{align*}
Further, in the above constant-current discharging scenario, the evolution of $V_1$ follows
\begin{align}\label{V1-equation}
V_1(t) = e^{-\beta_5t}V_1(0)-I\beta_4\left(1-e^{-\beta_5t}\right),
\end{align}
where
\begin{align*}
\beta_4 = R_1,  \ 
\beta_5  = \frac{1}{R_1C_1}.
\end{align*}
Since the battery has idled for a long  period prior to discharging, $V_1(0)$ relaxes at zero and  can be removed from~\eqref{V1-equation}. 

Then, combining~\eqref{measurement-equation},~\eqref{R0-SoC},~\eqref{final-Vs} and~\eqref{V1-equation}, the terminal voltage response is given by
\begin{align}\label{compact-terminal-voltage}\nonumber
V(\bm\theta;t) & = \sum_{i=0}^5 \alpha_i V_s^{i} (\bm\theta;t)
+I\theta_3\left(1-e^{-\theta_4t} \right)+I\theta_{5}\\
&\quad +I\theta_{6}e^{-\theta_{7}{\rm SoC}(t)}
+I\theta_{8}e^{-\theta_{9}\left(1-{\rm SoC}(t)\right)}.
\end{align}
with
\begin{align*}
{\bm \theta}= \left[\begin{matrix}  \beta_2 & \beta_3  & \beta_4 & \beta_5 & \gamma_1 & \gamma_2 & \gamma_3  &  \gamma_4 & \gamma_5\end{matrix}  \right]^{\T} ,
\end{align*}
and
\begin{align*}
V_s(\bm\theta;t) &= V_s(0)+{It}/Q_t+\theta_1I\left(1-e^{-\theta_2t}  \right),\\
{\rm SoC}(t) & = {\rm SoC}(0)+It/Q_t.
\end{align*} 

\subsubsection{Data-Fitting-Based Identification of $\bm \theta$}\label{Sec:Identification-Method}
In above, the terminal voltage $V$ is expressed in terms of $\bm\theta$, allowing one to identify $\bm\theta$ by minimizing the difference between the measured voltage and the voltage predicted by~\eqref{compact-terminal-voltage}. Hence, a data fitting problem similar to the one in Section~\ref{Sec:SoC-OCV-identification} can be formulated. It should be noted that  the resultant optimization will be nonlinear and nonconvex due to the presence of $h(\cdot)$. As a consequence, a numerical   algorithm   may get stuck in local minima and eventually give   unreasonable estimates. A  promising way of  mitigating this challenge is to constrain the numerical optimization search within  
a parameter space that is believably correct. Specifically, one can roughly determine the lower and upper bounds of part   or all of the
parameters,   set up a limited search space, and run numerical optimization within this space. With this notion, the identification problem can be formulated as a constrained optimization problem:
\begin{subequations}\label{Prediction-error-minimization}
\begin{align}\label{objective-function}
\hat {\bm \theta} =& \arg \min_{\bm \theta}  {1 \over 2} \left[\bm y-{\bm V}(\bm\theta)\right]^{\T} {\bm Q}^{-1}\left[\bm y-{\bm V}(\bm\theta)\right], \\ \label{constraint}
  \textrm{s.t.} & \  \underline{\bm\theta} \leq \bm \theta \leq \overline{\bm\theta },
\end{align}
\end{subequations}
where $\hat {\bm \theta}$ is the estimate of $\bm \theta$,  $\underline{\bm\theta}$ and $ \overline{\bm\theta }$ are the pre-set lower and upper bounds of $\bm\theta$, respectively, $\bm y$ the terminal  voltage measurement vector, $\bm Q$ an $M\times M$ symmetric positive definite matrix representing the covariance of the measurement noise, with $M$ being the number of the data points. 
Besides,  
\begin{align*}
{\bm y}  &= \begin{bmatrix} y(t_1) & y(t_2) & \cdots & y(t_M) \end{bmatrix}^{\T},\\
{\bm V}({\bm{\theta}})& = \begin{bmatrix} V(\bm \theta;t_1) & V(\bm \theta;t_2) & \cdots & V(\bm \theta;t_M) \end{bmatrix}^{\T}.
\end{align*}
Multiple numerical algorithms are available in the literature to solve~\eqref{Prediction-error-minimization}, a choice among which is the    interior-point-based trust-region method~\cite{tian2017parameter}. 

\begin{figure*}[t]
\captionsetup{justification = raggedright, singlelinecheck = false}
\centering
\includegraphics[trim = {10mm 40mm 20mm 20mm}, clip, width=0.85\textwidth]{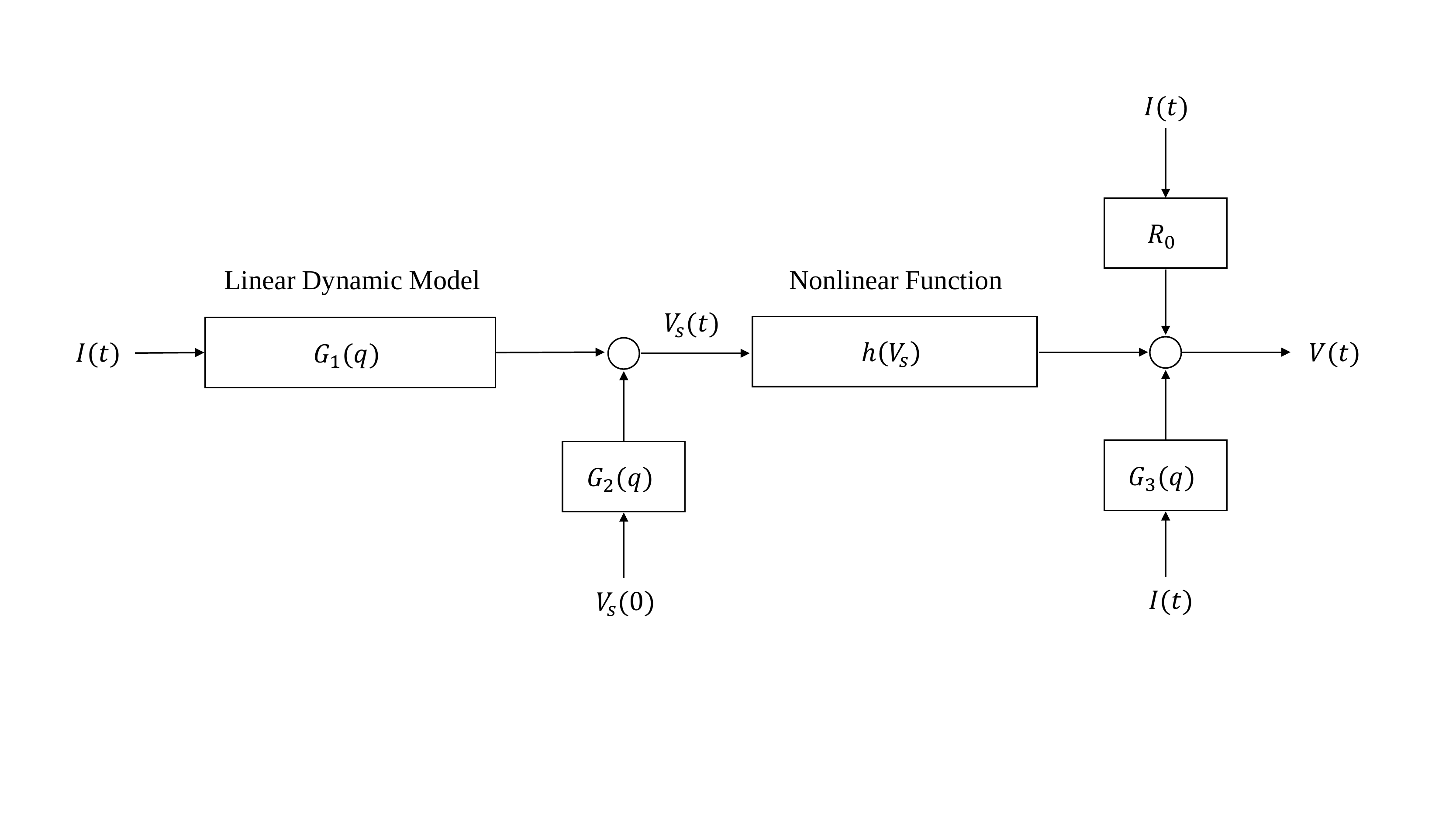}
\centering
\caption{The Wiener-type structure of the nonlinear double-capacitor (NDC) model.}
\label{Fig:Wiener-model-diagram}
\vspace{0mm}
\end{figure*}

\section{Parameter Identification  2.0: Variable-Current Charging/Discharging}\label{Sec:Parameter-Identification-2}

While it is not  unusual to charge or discharge a battery at a constant current, real-world battery systems such as those in electric vehicles generally operate at variable currents. Motivated by practical utility, an interesting and challenging question is: Will it be possible to estimate all the parameters of the NDC model in one shot when an almost arbitrary  current profile is applied to a battery? Having this question addressed will greatly improve the availability of the model, even to an on-demand level, for  battery management tasks. This section offers a  study in this regard from a Wiener identification perspective. It first unveils the NDC model's inherent Wiener-type structure and then develops an MAP-based identification approach. Here, the study assumes $R_0$ to be constant for convenience. 

\subsection{Wiener-Type Strucutre of the NDC Model}
The NDC model is structurally similar to a Wiener system---the   double RC circuits constitute a linear dynamic subsystem, and cascaded with it is a  nonlinear mapping. The following outlines   the discrete-time  Wiener-type formulation of~\eqref{state-space-equation}. 

Suppose that~\eqref{state-equation} is sampled with a time period $\Delta T$ and then discretized by the zero-order-hold (ZOH) method. The discrete-time model is expressed as
\begin{align}\label{discrete-time-equation}
x(t_{k+1})=A_dx(t_k)+B_dI(t_k), 
\end{align}
where $k$ is the discrete-time index with $t_k=k\Delta T$, and
\begin{align*}
A_d = e^{A\Delta T}, \
B_d = \left(\int^{\Delta T}_0e^{A\tau}{\rm d}\tau\right) B.
\end{align*}
Let us use $t$ instead of $t_k$ to represent the discrete time instant in sequel for notational simplicity. 
Then, \eqref{discrete-time-equation} can be written as
\begin{align*}
x(t) = (qI_{3\times 3}-A_d)^{-1}B_dI(t)+(qI_{3\times 3}-A_d)^{-1}qx(0),
\end{align*}
where $q$ is the forward shift operator, and $I_{3\times 3}\in\mathbb{R}^{3\times3}$ is an identity matrix, respectively. Since  $
V_s(t) = \begin{bmatrix} 0 & 1 & 0 \end{bmatrix} x(t)$ and $
V_1(t) = \begin{bmatrix} 0 & 0 & 1 \end{bmatrix} x(t)$, one can obtain the following after some lengthy derivation:
\begin{align}\label{Vs-transfer-function-original}
V_s(t) &= G_1(q)I(t)+G_2(q)V_s(0),\\  \label{equation-V1}
V_1(t) &=G_3(q)I(t)+G_4(q)V_1(0),
\end{align}
where
\begin{align*}
G_1(q) & = \frac{(\beta_1+\beta_2)q^{-1}-(\beta_1\beta_3+\beta_2)q^{-2}}{1-(1+\beta_3)q^{-1}+\beta_3q^{-2}},\\
G_2(q) & = \frac{1}{1-q^{-1}}, \\
G_3(q)&=\frac{\beta_4q^{-1}}{1+\beta_5q^{-1}},\\
G_4(q)&=\frac{1}{1+\beta_5q^{-1}},
\end{align*}
with 
\begin{align*} 
{\beta}_1 &= \frac{A_{21}B_{11}+A_{12}B_{21}}{A_{12}+A_{21}}\Delta{T},  \\
{\beta}_2  &=  \frac{A_{21}(B_{21}-B_{11})}{(A_{12}+A_{21})^2}\left(1-\beta_3\right),\\ 
\beta_3 &=e^{-(A_{12}+A_{21})\Delta{T}},\\
\beta_4 &= -\left(\beta_5 +1 \right)B_{31}/A_{33}, \\
\beta_5 &= -e^{A_{33}\Delta T}.
\end{align*}
Note that the notation $\beta$ is slightly abused above without causing confusion. Assume that  the battery has been at rest for a sufficiently long time to achieve an equilibrium state before a test. In this setting, $V_s(0)={\rm SoC}(0)$, $V_1(0)=0~\rm V$, and  $G_4(q)V_1(0)=0$. Besides, one can also see that the same parameter redundancy issue as in Section~\ref{Sec:Impedance-Identification} occurs again---only three parameters, $\beta_1$ through $\beta_3$, appear in~\eqref{Vs-transfer-function-original}, but four physical parameters, $C_b$, $C_s$, $R_b$ and $R_s$, need to be identified. To fix this, let $R_s=0$ as was done before. Then $\beta_1$ through $\beta_3$ reduce to be
\begin{align} \nonumber
{\beta}_1  = \frac{\Delta T}{C_b+C_s},\ 
{\beta}_2  =  \frac{R_bC_b^2\left(1-{\beta}_3 \right)}{(C_b+C_s)^2},\ 
{\beta}_3 = e^{-\frac{C_b+C_s}{C_bC_sR_b}\Delta{T}}.
\end{align}
If $\beta_1$ through $\beta_5$ become available,  the physical parameters can be reconstructed as follows:
\begin{align*}
C_b & = \frac{\Delta{T}}{{\beta}_1}-C_s,\
C_s  = \frac{ \left(1-{\beta}_3\right)\Delta T}{{\beta}_1-{\beta}_1{\beta}_3-{\beta}_2{\rm log} {\beta}_3},\\  
R_b & = -\frac{\left(\Delta T \right)^2}{C_bC_s{\beta}_1{\rm log} {\beta}_3},\
R_1  = \frac{-\beta_4}{\beta_5+1},\
C_1  = \frac{-\Delta T}{{\rm log}(-\beta_5)R_1}.
\end{align*}
Finally, it is obvious that 
\begin{align}\label{V-general-form}
V(t) = h\left[G_1(q)I(t)+G_2(q)V_s(0)\right]-G_3(q)I(t)+R_0I(t).
\end{align}
The above equation reveals the   block-oriented Wiener-type structure of  the NDC model, as depicted in Figure~\ref{Fig:Wiener-model-diagram}, in which the linear dynamic model $G_1(q)$ and the nonlinear function $h(V_s)$ are  interconnected sequentially. Given~\eqref{V-general-form}, the next pursuit is to estimate all of the  parameters simultaneously, which include $\alpha_i$ for $i=1,2,\dots,4$, ${\beta}_i$ for $i=1,2,\dots,5$, and $R_0$.  Here, $\alpha_0$ and $\alpha_5$ are free of identification as they can be expressed by $\alpha_i$ for $i=1,2,\dots,4$ (see Section~\ref{Sec:SoC-OCV-identification}). 

\subsection{MAP-Based Wiener Identification}

Consider the following model based on~\eqref{V-general-form} for notational convenience:
\begin{align}\label{V-compact-form}
z(t) &=  V(\bm\theta;u(t)) + v(t),
\end{align}
where $u$ is the input current $I$, $z$ the measured voltage, $v$   the measurement noise added to $V$ and assumed to follow a Gaussian distribution $\mathcal{N} (0, \sigma^2)$, and
\begin{align*}
 V(\bm\theta;u(t) ) 
& = h\left[G_1(q,\bm\theta)u(t)+G_2(q)V_s(0),\bm\theta\right]\\&\quad -G_3(q,\bm\theta)u(t) +\theta_{10}u(t),
\end{align*}
with
\begin{align}\nonumber
\bm\theta =\begin{bmatrix}\alpha_1& \alpha_2 &\alpha_3 &\alpha_4 &\beta_1& \beta_2 &\beta_3 & \beta_4&\beta_5 & R_0\end{bmatrix}^{\T}.
\end{align}
The input   and output datasets are denoted as 
\begin{align*}
{\bm u}  &= \begin{bmatrix} u(t_1) & u(t_2) & \cdots & u(t_N) \end{bmatrix}^{\T}\in\mathbb{R}^{N\times1},\\
{\bm z}  &= \begin{bmatrix} z(t_1) & z(t_2) & \cdots & z(t_N) \end{bmatrix}^{\T}\in\mathbb{R}^{N\times1},
\end{align*}
where $N$ is the total number of data samples. A combination of them is expressed as 
\begin{align*}
{\bm Z} = \begin{bmatrix}   {\bm u}& {\bm z}
\end{bmatrix}. 
\end{align*}
An ML-based   approach is developed in~\cite{Hagenblad:AUTO:2008} to deal with  Wiener system identification. If applied to~\eqref{V-compact-form}, it leads to consideration of the following problem:
\begin{align*}
\hat{\bm\theta}&= {\rm arg}\max_{\bm\theta}p(\bm Z | \bm\theta). 
\end{align*}
Following this line, one can derive a likelihood cost function and perform minimization to find out $\hat {\bm \theta}$. However, this method can be vulnerable to the risk of   local minima because of the nonconvexity issue resulting from the static nonlinear function $h(\cdot)$. This can cause unphysical   estimates. While carefully selecting an initial guess is  suggested   to alleviate this problem~\cite{Hagenblad:Thesis:1999}, it is often found  inadequate for many practical systems. In particular, our study showed that it could hardly deliver  reliable parameter estimation when used to handle the NDC model identification. 

MAP-based Wiener identification  thus is proposed here to overcome this problem. The MAP estimation can  incorporate  some prior knowledge about parameters to help drive the parameter search toward a reasonable minimum point. Specifically,   consider maximizing the {\em a posteriori} probability distribution of $\bm \theta$ conditioned on $\bm{Z}$:
\begin{align}\label{MAP}
\hat{\bm\theta}= {\rm arg}\max_{\bm\theta} p(\bm\theta|\bm Z). 
\end{align}
By the Bayes' theorem, it follows that
\begin{align}\nonumber
 p(\bm\theta|\bm Z)
 = \frac{p(\bm Z | \bm \theta) \cdot p(\bm \theta)}{p(\bm Z)}  \propto p(\bm Z|\bm \theta)\cdot p(\bm \theta).
\end{align}
In above, $p(\bm \theta)$ quantifies the  prior information available about $\bm \theta$. A general way is to characterize it as a Gaussian random vector following the distribution $p(\bm\theta) \sim \mathcal{N} \left(\bm m, {\bm P}\right)$.  Based on~\eqref{V-compact-form}, $p(\bm z|\bm\theta)\sim \mathcal{N} \left(\bm V(\bm\theta; \bm u ), {\bm R}\right)$, where  $\bm R = \sigma^2 \bm I$  and
\begin{align*}
\bm V(\bm\theta; \bm u )  &= \begin{bmatrix} V(\bm\theta;  u(t_1) )  & \cdots & V(\bm\theta;  u(t_N) )  \end{bmatrix}^{\T}.
\end{align*}
Then,
\begin{align*}
& p(\bm Z|\bm \theta) \cdot p(\bm \theta) \\
&\quad \propto  {\rm exp} \bigg(- \frac {1}{2}\left[\bm z- {\bm V} \left(\bm \theta; \bm u \right) \right]^{\T} \bm R^{-1}  \left[\bm z- {\bm V} \left(\bm \theta; \bm u\right) \right] \bigg)  \\ 
&  \quad \quad \   \cdot{\rm exp}\left( -\frac {1}{2} \left(\bm \theta-\bm m\right)^{\T} {\bm P}^{-1}  \left(\bm \theta-\bm m\right) \right).
\end{align*}
If using the  log-likelihood,  the problem  in~\eqref{MAP} is equivalent to
\begin{align}\label{MAP-cost-function}
\hat{\bm\theta}= {\rm arg}\min_{\bm\theta} J(\bm \theta),
\end{align}
where
\begin{align*}
J(\bm\theta)&=\frac {1}{2}\left[\bm z- {\bm V} \left(\bm \theta; \bm u \right) \right]^{\T} \bm R^{-1}   \left[\bm z- {\bm V} \left(\bm \theta; \bm u\right) \right] \\ &\quad +\frac{1}{2}\left(\bm\theta-\bm m \right)^{\T}\bm P^{-1}\left(\bm\theta-\bm m \right).
\end{align*}
For the nonlinear optimization problem in~\eqref{MAP-cost-function}, one can exploit the quasi-Newton method to numerically solve it~\cite{Hagenblad:AUTO:2008}. This method   iteratively updates the parameter estimate through
\begin{align}\label{theta-update}
\bm\theta_{k+1}=\bm\theta_{k}+\lambda_k \bm s_k.
\end{align}
Here, $\lambda_k$ denotes the step size at iteration step $k$, and $\bm s_k$ is the gradient-based search direction given by
\begin{align}\label{s_k-compute}
\bm s_k = -\bm B_k\bm g_k ,
\end{align}
where   $\bm B_k\in\mathbb{R}^{10\times10}$ is a positive definite matrix that approximates the Hessian matrix $\nabla^2 J\left( \bm\theta_{k}\right)$, and  $\bm g_k=\nabla J\left(\bm\theta_{k}\right)\in\mathbb{R}^{10\times1}$. Based on the well-known BFGS update strategy~\cite{wright1999numerical},  $\bm B_k$ can be updated by
\begin{align}\label{Bk-equation}
\bm B_k=\left(\bm I-\frac{\bm\delta_k\bm\gamma_k^{\T}}{\bm\delta_k^{\T}\bm\gamma_k}\right)\bm B_{k-1}\left(\bm I-\frac{\bm\gamma_k\bm\delta_k^{\T}}{\bm\delta_k^{\T}\bm\gamma_k}\right)+\frac{\bm\delta_k\bm\delta_k^{\T}}{\bm\delta_k^{\T}\bm\gamma_k},
\end{align}
with $\bm\delta_k = \bm\theta_k-\bm\theta_{k-1}$ and $\bm\gamma_k = \bm g_k-\bm g_{k-1}$. In addition,   
\begin{align}\nonumber \label{g_k-update}
\bm g_k &= - \left( \frac{\partial{\bm V}\left(\bm\theta_{k}; \bm u\right)}{\partial\bm\theta_{k}} \right)^{\T} \bm R^{-1}
\left[\bm z-\bm V\left( \bm\theta_{k} ; \bm u \right) \right]\\ &\quad +\bm{P}^{-1}\left(\bm\theta_k-\bm{m} \right),
\end{align}
where each column of $\frac{\partial{\bm V}\left(\bm\theta; \bm u\right)}{\partial\bm\theta}\in\mathbb{R}^{N\times 10}$ is given by
\begin{align} \nonumber
\frac{\partial{\bm V}\left(\bm\theta; \bm u\right)}{\partial\theta_{i}} &= \bm x^{\circ i}-\bm x^{\circ 5}~{\rm for}~i = 1,2,\dots,4, \\ \nonumber
\frac{\partial{\bm V}\left(\bm\theta; \bm u\right)}{\partial\theta_{5}}&=\bm\Sigma\circ \frac{q^{-1}-\theta_7q^{-2}}{1-(1+\theta_7)q^{-1}+\theta_7q^{-2}}\bm u, \\ \nonumber
\frac{\partial{\bm V}\left(\bm\theta; \bm u\right)}{\partial\theta_{6}}&=\bm\Sigma\circ \frac{q^{-1}-q^{-2}}{1-(1+\theta_7)q^{-1}+\theta_7q^{-2}}\bm u, \\ \nonumber
\frac{\partial{\bm V}\left(\bm\theta; \bm u\right)}{\partial\theta_{7}}&=\bm\Sigma\circ \frac{\theta_6q^{-2}-2\theta_6q^{-3}+\theta_6q^{-4}}{\left(1-(1+\theta_7)q^{-1}+\theta_7q^{-2}\right)^2}\bm u,  \\ \nonumber
\frac{\partial{\bm V}\left(\bm\theta; \bm u\right)}{\partial\theta_{8}}& = \frac{-q^{-1}}{1+\theta_9q^{-1}}\bm u,\\ \nonumber
\frac{\partial{\bm V}\left(\bm\theta; \bm u\right)}{\partial\theta_{9}}& = \frac{\theta_{8}q^{-2}}{1+2\theta_9q^{-1}+\theta_9^2q^{-2}}\bm u,\\ \nonumber
\frac{\partial{\bm V}\left(\bm\theta; \bm u\right)}{\partial\theta_{10}}& = \bm u,
\end{align}
with 
\begin{align}\nonumber
\bm{x}& = G_1(q,\bm\theta)\bm u+G_2(q)V_s(0)\bm{1},\\ \nonumber
\bm\Sigma &=\sum^4_{i=1}i\theta_i{\bm x}^{\circ(i-1)}+5\left(\overline{V}-\underline{V}-\sum^4_{i=1}\theta_i\right)\bm x^{\circ 4}.
\end{align}
Here, $\bm x\circ \bm u$ denotes the Hadamard product of $\bm x$ and $\bm u$, $\bm x^{\circ2}$ denotes the Hadamard power with $\bm x^{\circ2}=\bm x\circ \bm x$, and $\bm1\in\mathbb{R}^{N\times1}$ denotes a column vector with all elements equal to one. 

 Finally, note that $\lambda_k$ needs to be chosen carefully to make $J(\bm \theta)$ decrease monotonically. One can use the Wolfe conditions and let $\lambda_k$ be selected such that
\begin{subequations}\label{Wolfe-conditions}
\begin{align}
{J}\left(\bm\theta_k+\lambda_k \bm s_k \right) & \leq {J}\left(\bm\theta_k\right)+c_1\lambda_k \bm g_k^{\T}\bm s_k,\\
\nabla J\left(\bm\theta_{k}+\lambda_k\bm{s}_k\right)^{\T}\bm{s}_k& \geq c_2 \nabla J\left(\bm\theta_{k}\right)^{\T}\bm{s}_k,
\end{align}
\end{subequations}
with $0<c_1<c_2<1$. For the quasi-Newton method, $c_1$ is usually set to be quite small, e.g., $c_1=10^{-6}$, and $c_2$ is typically set to be 0.9. The selection of $\lambda_k$ can be based on trial and error in implementation. One can start   with  picking a  number and check  the Wolfe conditions. If the conditions are not satisfied, reduce the number and check again. An interested reader is referred to~\cite{wright1999numerical} for detailed discussion  about the $\lambda_k$ selection. Summarizing the above,  Table~\ref{Table:Quasi-Newton-method} outlines the implementation procedure for the MAP-based Wiener identification.  

\begin{table}[t]
\captionsetup{justification = raggedright, singlelinecheck = false}
\caption{Quasi-Newton-based implementation for MAP-based Wiener identification.}
\centering
\framebox[\linewidth]{
\begin{minipage}{0.97\linewidth}\normalsize
\begin{algorithmic} 
\STATE Initialize $\bm\theta_0$ and set the convergence tolerance 
\REPEAT 
\STATE{ Compute   $\bm g_k$ {\em via}~\eqref{g_k-update}
\IF {$k=0$}
\STATE Initialize $\bm B_0=0.001\frac{1}{\lVert{\bm g_0}\rVert}\bm I$ 
\ELSE
\STATE Compute $\bm B_k$ {\em via}~\eqref{Bk-equation}
\ENDIF

\STATE Compute   $\bm{s}_k$ {\em via}~\eqref{s_k-compute}
\STATE Find  $\lambda_k$ that satisfies the Wolfe conditions~\eqref{Wolfe-conditions}


\STATE Perform the update {\em via}~\eqref{theta-update}

}
\UNTIL{${J}(\bm\theta_k)$ converges}

\RETURN {$\hat{\bm\theta}=\bm\theta_k$} 
\end{algorithmic}
\end{minipage}
}
\label{Table:Quasi-Newton-method}
\end{table}

\begin{remark}
 While the MAP estimation has enjoyed  a long history of addressing a   variety of estimation problems, no study has been reported about its application to Wiener system identification to our knowledge. Here, it is found to be a very useful approach for providing physically reasonable parameter estimation for practical systems, as  it takes into account some prior knowledge about the unknown parameters. In a Gaussian setting as adopted here, the prior $p(\bm \theta)$ translates into a regularization term in $J(\bm \theta)$, which prevents incorrect fitting and enhances the robustness of the numerical optimization against nonconvexity. 
 
\end{remark}
\begin{remark}
 The proposed 2.0 identification approach requires some prior knowledge of the parameters to be available, which can be developed in several ways in practice. First, $R_0$ can be roughly estimated using the voltage drop at the beginning of the discharge, to which it is a main contributor. Second, the polynomial coefficients of $h(\cdot)$ can be approximately obtained from an experimentally calibrated SoC-OCV curve if there is any. Third, one can derive a rough range for $C_b+C_s$ if a battery's capacity is approximately known. Finally, as the parameters of batteries of the same kind and brand are usually close, one can take the parameter estimates acquired from one battery as prior knowledge for another. 
\end{remark}

\begin{remark}
In general, a prerequisite for  successful identification is that the parameters must be identifiable in a certain sense. Following along similar lines as in~\cite{Fang:CEP:2014,VanDoren:IFAC:2009}, one can  rigorously define the parameters' local identifiability for the considered Wiener identification problem and find out that  a sufficient condition for  it to hold is the full rankness of the sensitivity matrix ${\partial{\bm V}\left(\bm\theta; \bm u\right)}/{\partial\bm\theta}$, which can be used for identifiability testing. Using this idea, our simulations consistently showed the full rankness of the sensitivity matrix under variable current profiles such as those in Figure~\ref{Fig:varying-03}, indicating that the NDC model can be locally identifiable. Related with identification is  optimal input design, which concerns  designing the best current profile to maximize the parameter  identifiability~\cite{Rothenberger:JES:2015,Park:ACC:2018}. It will be part of our future research to explore this interesting problem for the NDC model.

\end{remark}

\begin{remark}
It is worth mentioning that the 2.0 identification approach can be readily extended to identify some other ECMs that have a Wiener-like structure like the Rint and Thevenin models.  One can follow similar lines to develop  the  computational procedures for each, and hence the details are skipped here.
\end{remark}

\begin{remark}
The 1.0 and 2.0 identification approaches are designed to perform offline identification for the NDC model, each with its own advantages. The 1.0 approach is designed for in-lab battery modeling and analysis, using simple two-step (trickle- and constant-current discharging) battery testing protocols. While requiring a long  time for experiments, it can offer high accuracy in parameter estimation. More sophisticated by design, the 2.0 approach can  extract the parameters all at once from data based on variable current profiles. It can be conveniently exploited to determine the NDC model for batteries operating in real-world applications.

\end{remark}


\section{Experimental Validation}\label{Sec:Experimental-Validation}
\begin{figure}[t]
\captionsetup{justification = raggedright, singlelinecheck = false}
\centering
\includegraphics[trim = {0mm 5mm 0mm 0mm}, clip, width=0.33\textwidth]{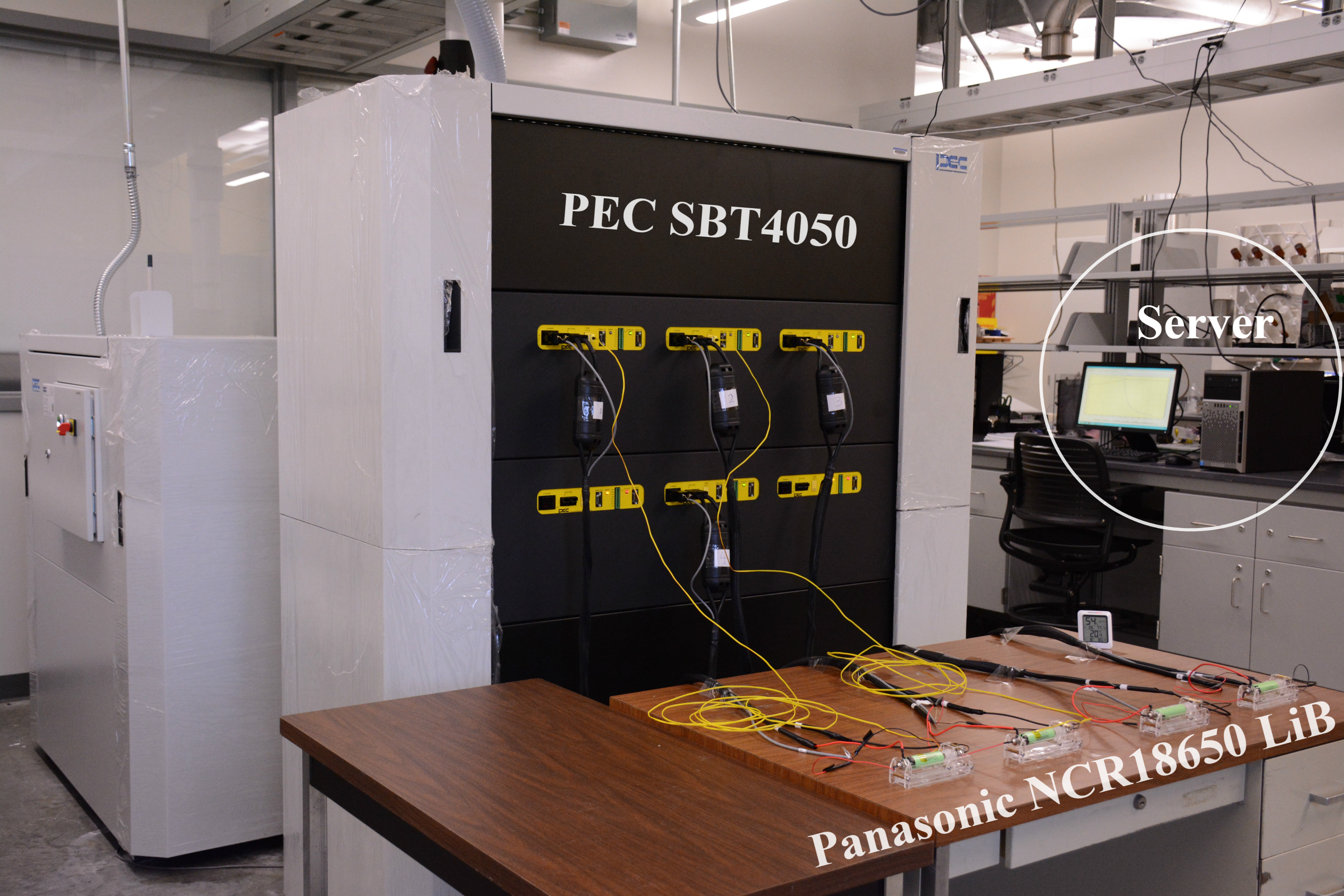}
\caption{PEC\textsuperscript{\textregistered} SBT4050 battery tester.}
\label{Fig:Experiment-System}
\vspace{0mm}
\end{figure}

This section presents experimental validation of the proposed NDC model and parameter identification 1.0 and 2.0 approaches. All the experiments in this section were conducted on a PEC\textsuperscript{\textregistered} SBT4050 battery tester (see Figure~\ref{Fig:Experiment-System}). It can support charging/discharging with arbitrary current-, voltage- and power-based loads (up to 40 V and 50 A). A specialized server  is used to prepare and configure a test offline and collect experimental data online {\em via} the associated software   LifeTest\textsuperscript{TM}. Using this facility, charging/discharging tests were performed to generate data on a Panasonic NCR18650B lithium-ion battery cell, which was set to operate between 3.2 V (fully  discharged) and 4.2 V (fully charged).

\subsection{Validation Based on Parameter Identification 1.0}\label{Sec:Validation-1.0}

This validation first  extracts the NDC model from  training dataset using the 1.0 identification approach in Section~\ref{Sec:Parameter-Identification-1} and then applies the identified model to   validation datasets to assess its predictive capability.



\begin{figure}[t]\vspace{-4mm}
\captionsetup{justification = raggedright, singlelinecheck = false}
\centering
\includegraphics[trim = {0mm 5mm 0mm 0mm}, clip, width=0.4\textwidth]{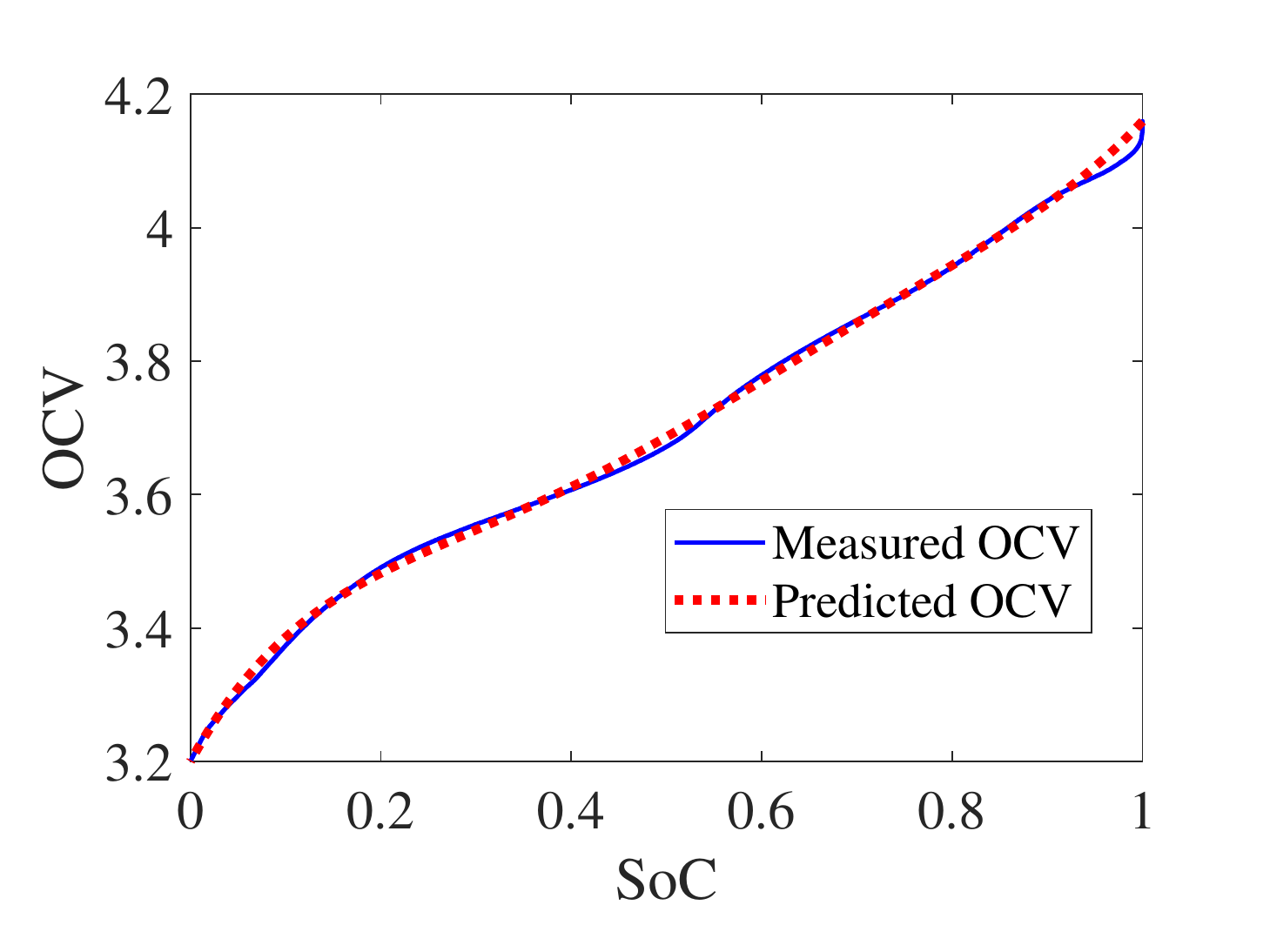}
\caption{Identification 1.0: parameter identification of $h(\cdot)$ that defines SoC-OCV relation.}
\label{Fig:SoC-OCV-Validation}
\vspace{0mm}
\end{figure}
\begin{figure}[t]
\captionsetup{justification = raggedright, singlelinecheck = false}
\centering
\includegraphics[trim = {0mm 5mm 4mm 9mm}, clip, width=0.4\textwidth]{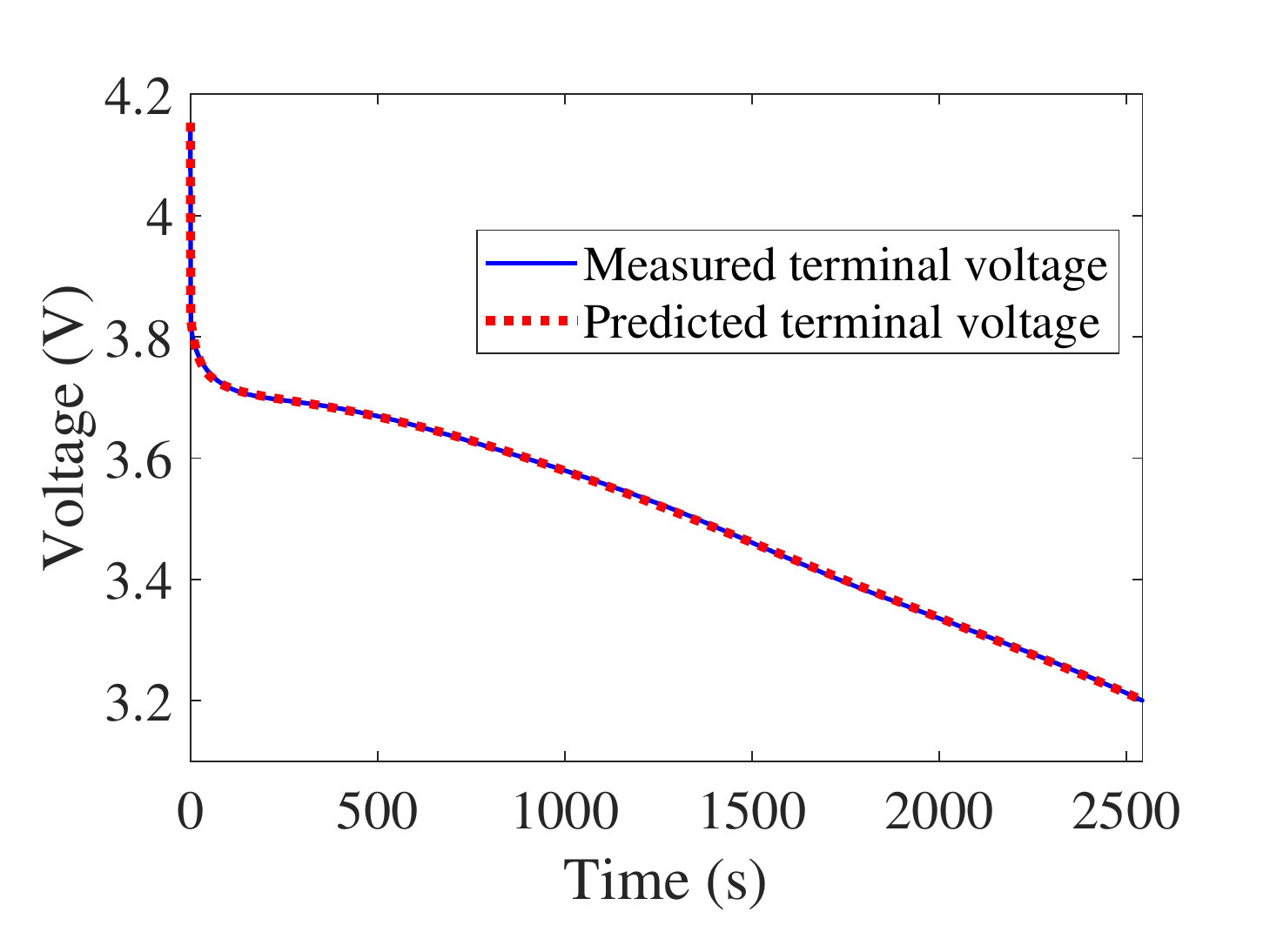}
\caption{Identification 1.0: model fitting with the training data obtained under 3 A constant-current discharging.} 
\label{Fig:3A-fitting}
\vspace{0mm}
\end{figure}

As a first step, the cell was fully charged and relaxed for a long time period. Then, a full discharge test was applied to the cell using a trickle constant current of 0.1 A (about 1/30 C-rate). With this test, the total capacity is determined to be $Q_t=3.06~{\rm Ah}$ by coulomb counting, implying $C_b+C_s=11,011~\rm F$. Further, from the SoC-OCV curve fitting, we obtain
\begin{align}\label{Theoretical-SoC-OCV} \nonumber
{\rm OCV}= & ~3.2+2.59\cdot{\rm SoC} -9.003\cdot{\rm SoC}^2 +18.87\cdot{\rm SoC}^3\\ \nonumber
& -17.82\cdot{\rm SoC}^4+6.325\cdot{\rm SoC}^5,
\end{align} 
which establishes $h(\cdot)$ immediately. The measured and identified SoC-OCV curves are compared in Figure~\ref{Fig:SoC-OCV-Validation}. Next, the cell was fully charged again and left idling for a long time. This was then followed by a full discharge using a constant current of 3 A to produce data for estimation of the impedance and capacitance parameters. The identification was achieved by solving the constrained optimization problem in~\eqref{Prediction-error-minimization}. The computation took around 1 sec, performed on a Dell Precision Tower 3620  equipped with 3 GHz Inter Xeon CPU, 16 Gb RAM and MATLAB R2018b. 
Table~\ref{Tab:Parameter-Setting} summarizes the initial guess, lower and upper bounds, and obtained estimates of the parameters. The    physical parameter estimates are extracted as: $C_b=10,037~{\rm F}$, $C_s=973~{\rm F}$, $R_b=0.019~\rm{\Omega}$, $R_s=0$,  $R_1=0.02~\rm{\Omega}$, $C_1=3,250~\rm F$, and 
\begin{align*}
R_0=0.0531+0.1077e^{-3.807\cdot{\rm SoC}}+0.0533e^{-7.613\cdot{\rm (1-SoC)}}.
\end{align*}
The model is now fully available from the two steps.  Figure~\ref{Fig:3A-fitting} shows that it accurately fits with the measurement data. 

\begin{table*}[t]\centering
{%
\begin{threeparttable}
\caption{Identification 1.0: initial guess, bound limits and identification results.}
\begin{tabular}{ccccccccccccccc}
\toprule%
Name&  $\beta_2$  & $\beta_3$ & $\beta_4$ & $\beta_5$   & $\gamma_1$ &$\gamma_2$ &$\gamma_3$ &$\gamma_4$ &$\gamma_5$ \\
\midrule
 Initial guess  &   0.02  &  0.05 &   0.005 &  1/100  &  0.05  &  0.2 & 8  & 0.07  &  12    \\
${\underline{\bm\theta}}$  &  0.005&      0.005 &       0.001  &     1/800 &   0.01  &   0.05&       1  &    0.01&    1  \\ 
${\overline{\bm\theta}}$  &  0.2 &     0.2 &     0.03  &      1/10   &   0.09 &   0.35 &   15  &    0.12 &   15    \\ 
$\hat{\bm\theta}$   &  0.0163  &  0.0575& 0.02 &  1/65   & 0.0531 & 0.1077  & 3.807&  0.0533 & 7.613 \\
\bottomrule
\end{tabular}
\begin{tablenotes}
\small
\item Note: quantities are given in  SI standard units in Tables II and III.
\end{tablenotes}
\label{Tab:Parameter-Setting}
\end{threeparttable}}
\vspace{2mm}
\end{table*}


\begin{figure}[tp!]
\captionsetup{justification = raggedright, singlelinecheck = false}
\centering
\includegraphics[trim = {0mm 5mm 4mm 0mm}, clip, width=0.4\textwidth]{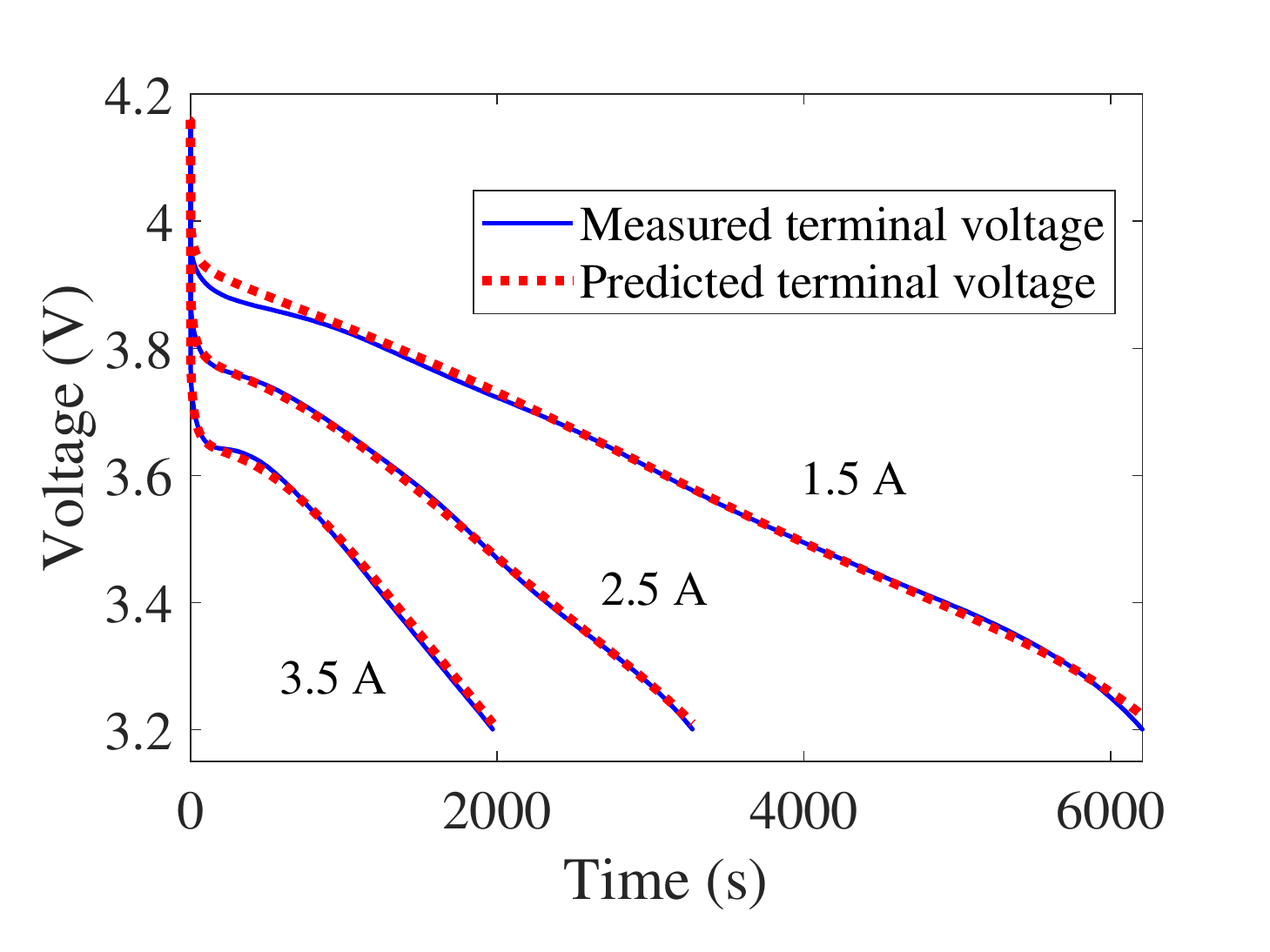}\centering
\caption{Identification 1.0: predictive fitting over   validation data  obtained by discharging at different constant currents.} 
\label{Fig:Fitting-Curve-123}
\vspace{0mm}
\end{figure}

\begin{figure*}[t]
\captionsetup{justification = raggedright, singlelinecheck = false}
\centering
\subfigure[]
{\includegraphics[trim = {40mm 2mm 30mm 18mm}, clip, width=0.75\textwidth]{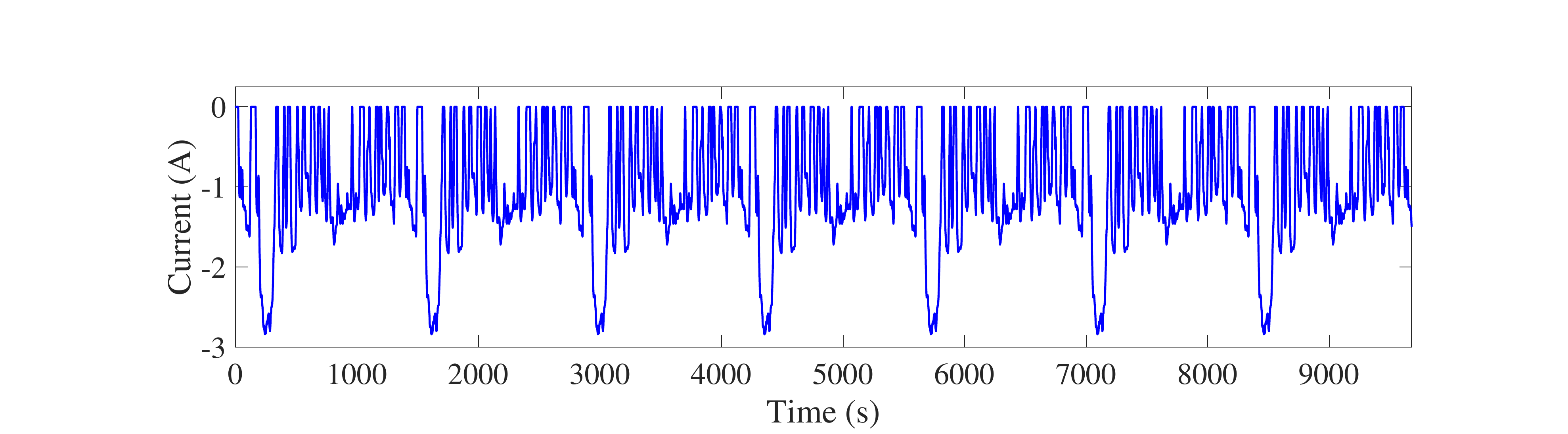}\label{Fig:varying-current-03}}
\subfigure[]
{\includegraphics[trim = {40mm 2mm 30mm 18mm}, clip, width=0.75\textwidth]{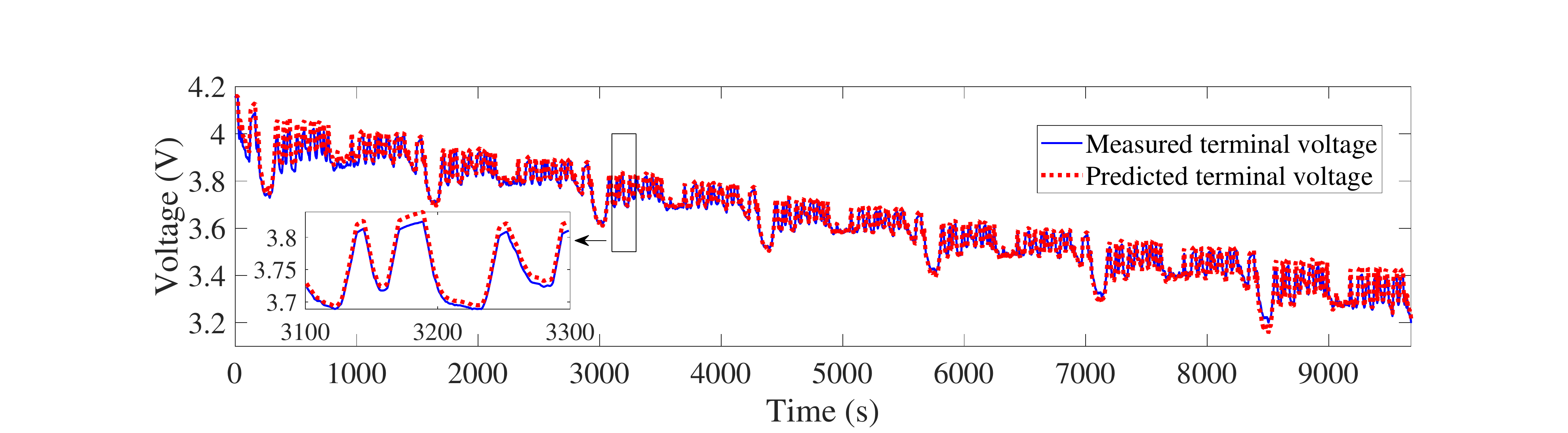}\label{Fig:varying-voltage-03}}
\caption{Identification 1.0: predictive fitting over   validation dataset obtained by discharging at varying currents (0$\sim$3 A). (a) Current profile. (b) Voltage fitting.} 
\label{Fig:varying-03}
\vspace{1mm}
\end{figure*}

\begin{figure*}[!htbp]
\captionsetup{justification = raggedright, singlelinecheck = false}
\centering
\subfigure[]
{\includegraphics[trim = {40mm 2mm 30mm 18mm}, clip, width=0.75\textwidth]{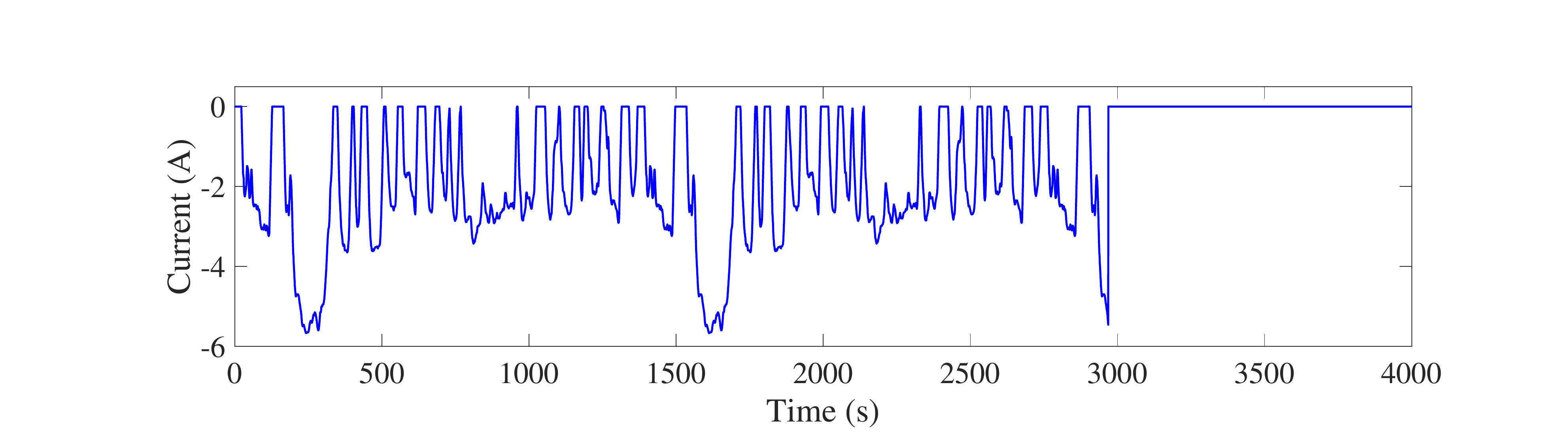}\label{Fig:varying-current-06}}
\subfigure[]
{\includegraphics[trim = {40mm 2mm 30mm 18mm}, clip, width=0.75\textwidth]{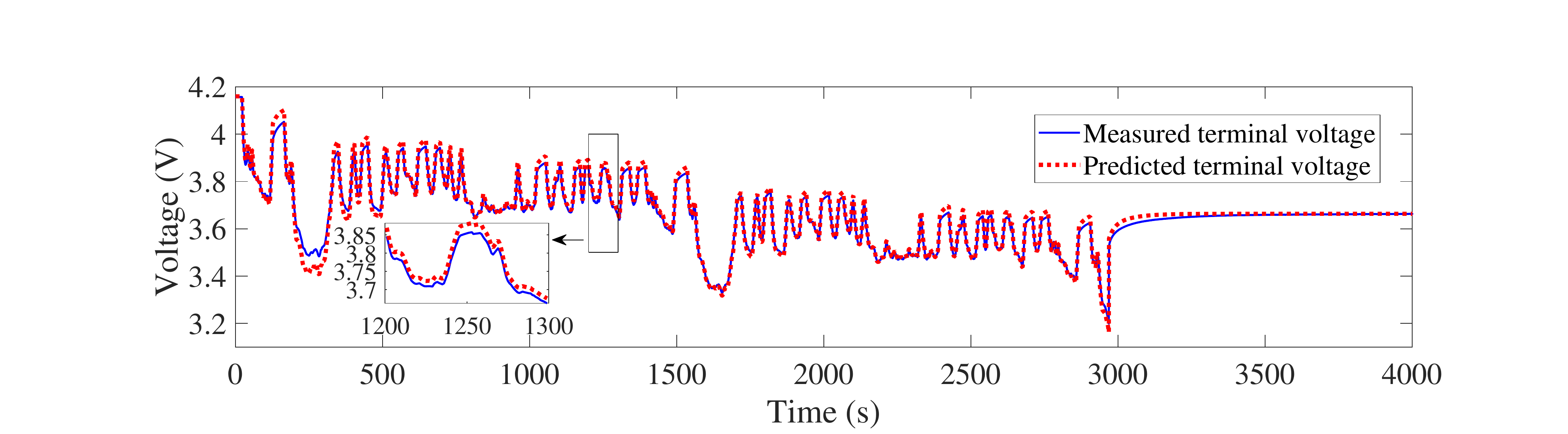}\label{Fig:varying-voltage-06}}
\caption{Identification 1.0: predictive fitting over   validation dataset  obtained by discharging at varying currents   ($0\sim6$ A). (a) Current profile. (b) Voltage fitting.} 
\label{Fig:varying-06}
\vspace{1mm}
\end{figure*}

%

While an identified  model generally can well fit a training dataset, it is more meaningful and revealing  to examine  its predictive performance  on some   different datasets. Hence, five more tests were conducted by discharging the cell using constant currents of 1.5 A, 2.5 A and 3.5 A and   two variable current profiles, respectively.  Figure~\ref{Fig:Fitting-Curve-123} shows what the  identified model predicts for discharging at constant currents. An overall high accuracy is observed, even though the prediction is slightly less accurate when the current is 1.5 A, probably because the parameters are current-dependent to a certain extent.  The variable current profiles  are portrayed in Figures~\ref{Fig:varying-current-03} and~\ref{Fig:varying-current-06}, which were created by scaling the Urban Dynamometer Driving Schedule (UDDS) profile in~\cite{UDDS}  to span  the ranges of $0$$\sim$$3$ A and  $0$$\sim$$6$ A, respectively.  Figures~\ref{Fig:varying-voltage-03} and~\ref{Fig:varying-voltage-06} present the predictive fitting results. Both of them illustrate that the model-based voltage prediction is quite close to the actual measurements. These results demonstrate the excellent predictive capability  of the NDC model. 

\subsection{Validation Based on Parameter Identification 2.0}

\begin{table*}[t]\centering
\begin{threeparttable}
\caption{Identification 2.0: initial guess, prior knowledge and identification results.}
\begin{tabular}{ccccccccccccc}
\toprule%
Name& $\alpha_1$ & $\alpha_2$ & $\alpha_3$& $\alpha_4$  & $\breve\beta_1$ &$\breve\beta_2$ &$\breve\beta_3$& $\beta_4$ & $\beta_5$    & $R_0$    \\
\midrule
Initial guess & 2.59 &   -9.003 &  18.87  & -17.82 & $9.078\times10^{-5}$ &  $8.914\times10^{-4}$ &  0.964 &$-4.938\times10^{-4}$ &$-0.9753$  &   0.08  \\ 
$\bm m$ &  - &  -  & -   &  -   &$9.078\times10^{-5}$& $8.914\times10^{-4}$ &  0.964&$-4.938\times10^{-4}$ &$-0.9753$ &  0.08 \\
$\sqrt{{\rm diag}(\bm P)}$&  - &  -  & -   &  - & $0.001\times m_5$ & $0.15\times m_6$ &  $0.15\times m_7$ &$0.15\times m_8$ &$0.15\times m_9$ &  $0.15\times m_{10}$ \\
$\hat{\bm \theta}$ &  2.32 & -8.15  & 19.345   &  -20.78& $9.082\times10^{-5}$ & $9.227\times10^{-4}$ &  0.982 &$-4.859\times10^{-4}$ &$-0.8153$  & 0.069 \\
\bottomrule
\end{tabular}
\label{Tab:Experimental-Results}
\end{threeparttable}
\vspace{1mm}
\end{table*}

Let us now consider the 2.0 identification approach developed in Section~\ref{Sec:Parameter-Identification-1}, which  treats the NDC model as a Wiener-type system and  performs MAP-based parameter estimation. This approach advantageously  allows all the parameters to be estimated in  a convenient one-shot procedure. 

Following the manner in Section~\ref{Sec:Validation-1.0}, one can apply the 2.0   approach   to a training dataset to extract an NDC model and then use it to predict the responses over several other different   datasets. The validation here is also set to   evaluate the NDC model  against  the Rint model~\cite{He:Energies:2011} and the Thevenin model with one serial RC circuit~\cite{He:Energies:2011}, which are commonly used in the literature. The comparison also extends to   a basic version of the NDC model (referred to as ``basic NDC'' in sequel), one with a constant $R_0$ and  without $R_1$-$C_1$ circuit, with the purpose   of examining the utility of the NDC model when it is reduced to a simpler form. Note that, even though the NDC model is the most sophisticated among them, all of the  four models offer high computational efficiency by requiring only a small number of  arithmetic operations.
 
 \begin{figure*}[t]
\captionsetup{justification = raggedright, singlelinecheck = false}
\centering
\subfigure[]
{\includegraphics[trim = {40mm 2mm 30mm 18mm}, clip, width=0.75\textwidth]{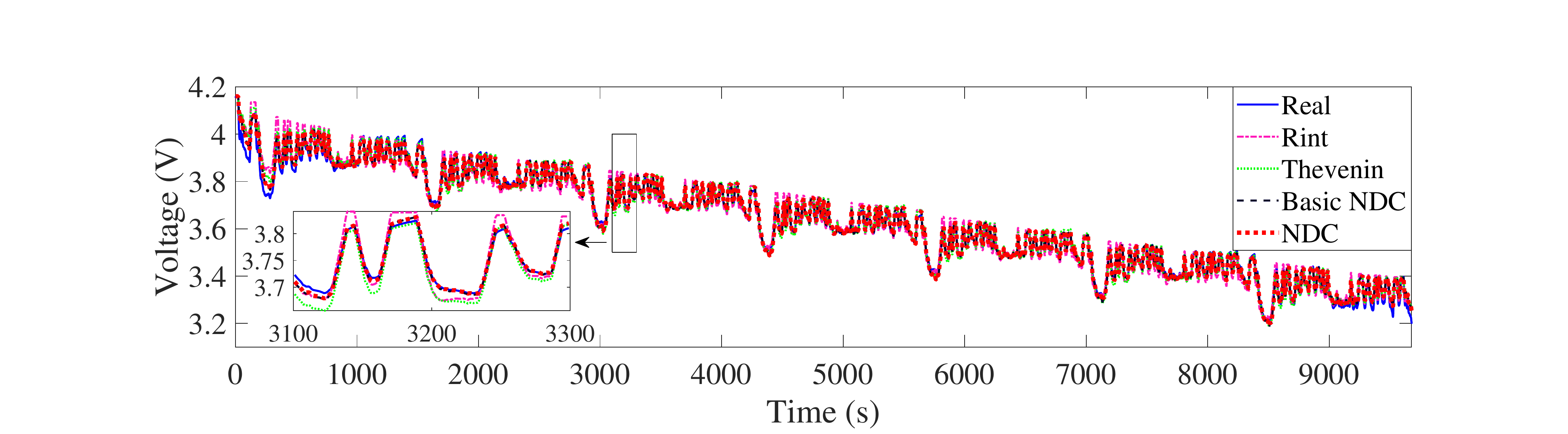}\label{Fig:varying-03-multimodel-voltage}}
\subfigure[]
{\includegraphics[trim = {40mm 2mm 30mm 18mm}, clip, width=0.75\textwidth]{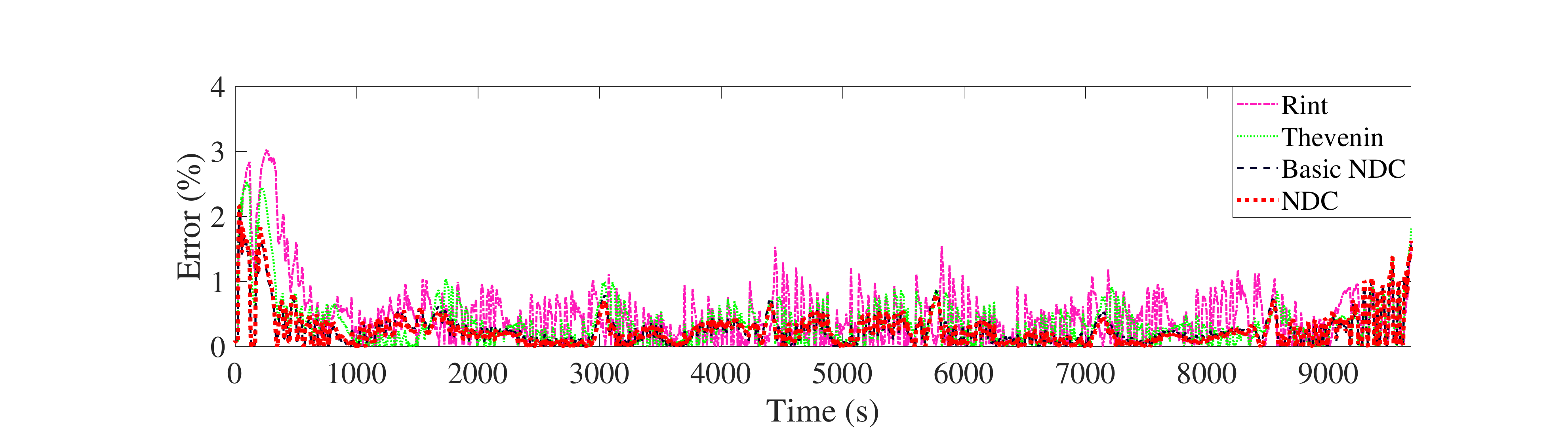}\label{Fig:varying-03-multimodel-error}}
\caption{Identification 2.0. (a) Model fitting with training dataset. (b) Fitting error in percentage.} 
\label{Fig:varying-03-multimodel}
\vspace{0mm}
\end{figure*}

\vspace{-2mm}
\begin{figure}[t]
\captionsetup{justification = raggedright, singlelinecheck = false}
\centering
\includegraphics[trim = {0mm 5mm 4mm 0mm}, clip, width=0.4\textwidth]{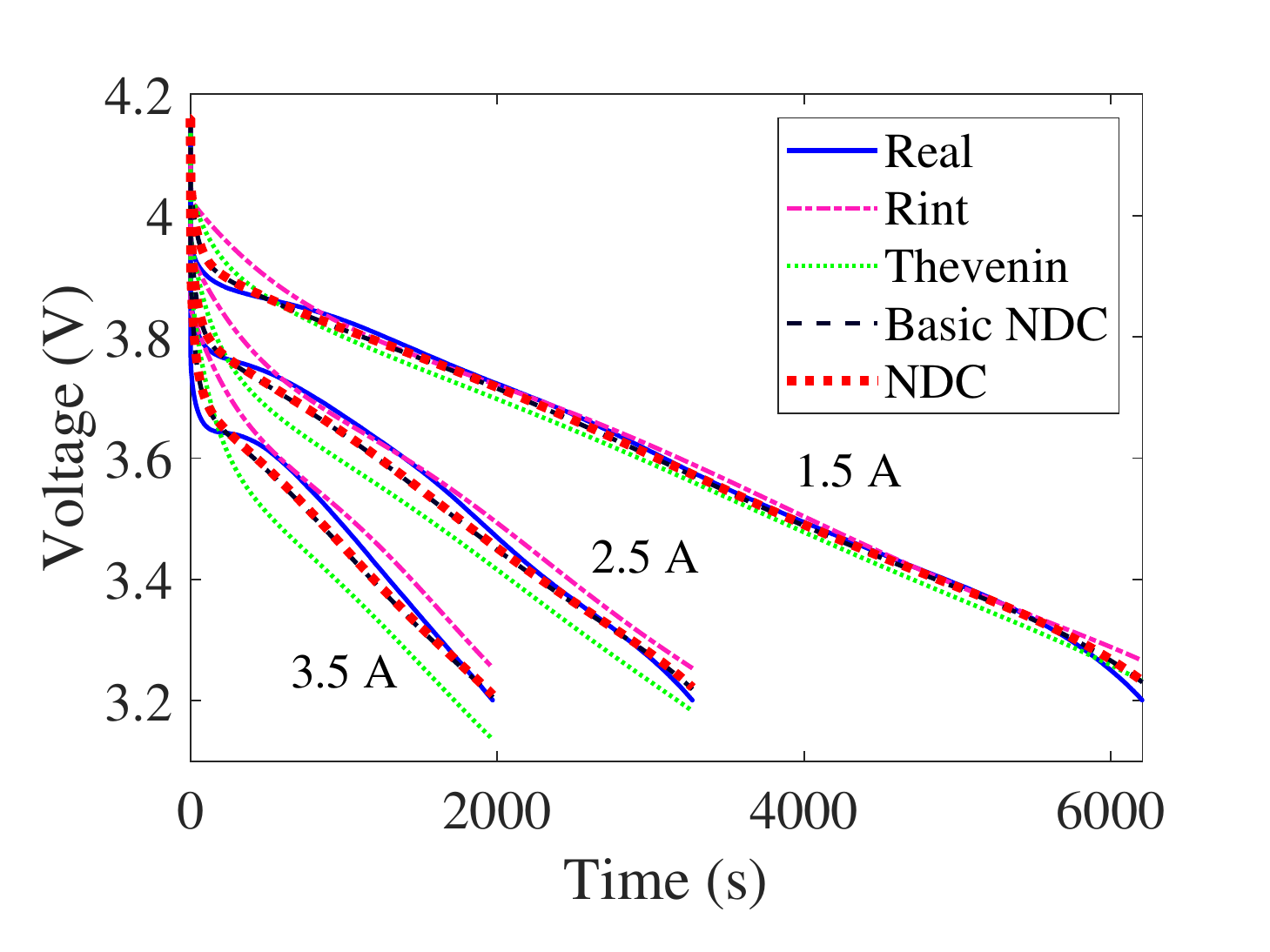}\centering
\caption{Identification 2.0: predictive fitting over   validation datasets  obtained by discharging at different varying currents.} 
\label{Fig:Fitting-Curve-123-multimodel}
\vspace{0mm}
\end{figure}
 

These four models are all Wiener-type, so the 2.0 identification approach can be used to identify them on   the same training dataset, i.e.,  the one shown in Figure~\ref{Fig:varying-03}, thus ensuring  a fair comparison. The parameter setting for   the NDC model identification and the estimation result are summarized in Table~\ref{Tab:Experimental-Results}. The computation took around 4 sec. The resultant physical parameter estimates are given by: $C_b=10,031~{\rm F}$, $C_s=979~{\rm F}$, $R_b=0.063~\rm{\Omega}$, $R_s=0$, $R_1=0.003~\rm{\Omega}$, $C_1=2,449~\rm F$ and $R_0 = 0.069~\rm\Omega$. The identification results for the Rint, Thevenin model and  basic NDC models are omitted here for the sake of space.

Figure~\ref{Fig:varying-03-multimodel-voltage} depicts how the identified  models fit with the training dataset. One can observe that the NDC model and its basic version show excellent fitting accuracy,   overall better than the Rint and Thevenin models. A more detailed comparison is given in Figure~\ref{Fig:varying-03-multimodel-error}, which displays the  fitting error in percentage. It is  seen that the Rint model shows the least  accuracy, followed by the Thevenin model. The NDC model and its basic version well outperform them, with the NDC model  performing slightly better. 

\begin{figure*}[t]
\captionsetup{justification = raggedright, singlelinecheck = false}
\centering
\subfigure[]
{\includegraphics[trim = {40mm 2mm 30mm 18mm}, clip, width=0.75\textwidth]{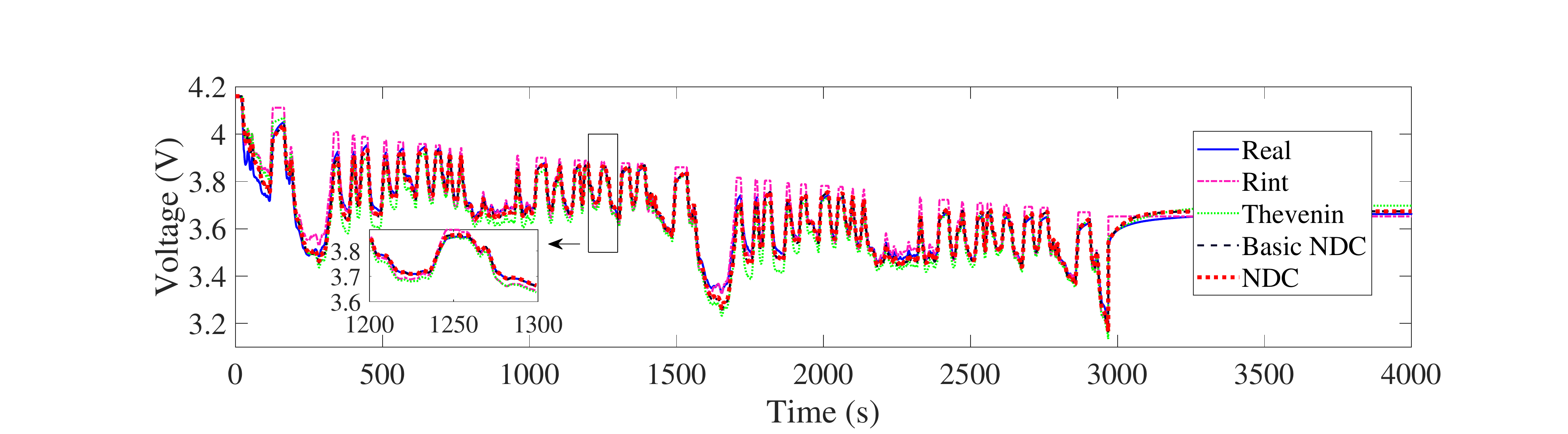}\label{Fig:varying-06-multimodel-voltage}}
\subfigure[]
{\includegraphics[trim = {40mm 2mm 30mm 18mm}, clip, width=0.75\textwidth]{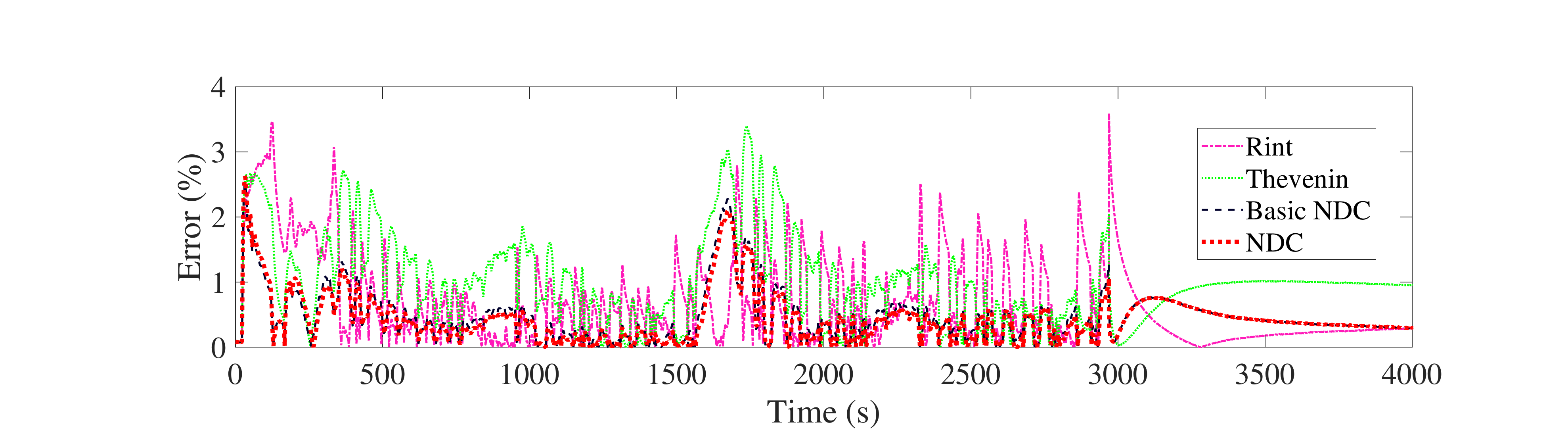}\label{Fig:varying-06-multimodel-error}}
\caption{Identification 2.0. (a) Predictive fitting over   validation dataset  obtained by discharging at varying currents   between 0 A and 6 A. (b) Predictive fitting error in percentage.} 
\label{Fig:varying-06-multimodel}
\vspace{0mm}
\end{figure*}

Proceeding forward, let us investigate the predictive performance of the four models over several validation datasets. First, consider the datasets obtained by constant-current discharging at  1.5 A, 2.5 A and 3.5 A, as illustrated in Figure~\ref{Fig:Fitting-Curve-123}. Figure~\ref{Fig:Fitting-Curve-123-multimodel} demonstrates that the NDC model and its basic version can predict the voltage responses under different currents much more accurately than the Rint and Thevenin models. Next, consider the dataset in Figure~\ref{Fig:varying-06} based on variable-current discharging. Figure~\ref{Fig:varying-06-multimodel} shows that the prediction accuracy of all the models is lower than the fitting accuracy, which is understandable. However, the NDC model and its basic version  are still again the most capable of predicting, with the   error mostly lying below $1\%$. As a contrast, while the Thevenin model can   offer a decent fit with the training dataset as shown in Figure~\ref{Fig:varying-03-multimodel}, its prediction accuracy over the validation dataset is not as satisfactory. This implies that  it is less predictive than the NDC model. 

\begin{figure}[t]
\captionsetup{justification = raggedright, singlelinecheck = false}
\centering
\includegraphics[trim = {0mm 5mm 0mm 0mm}, clip, width=0.4\textwidth]{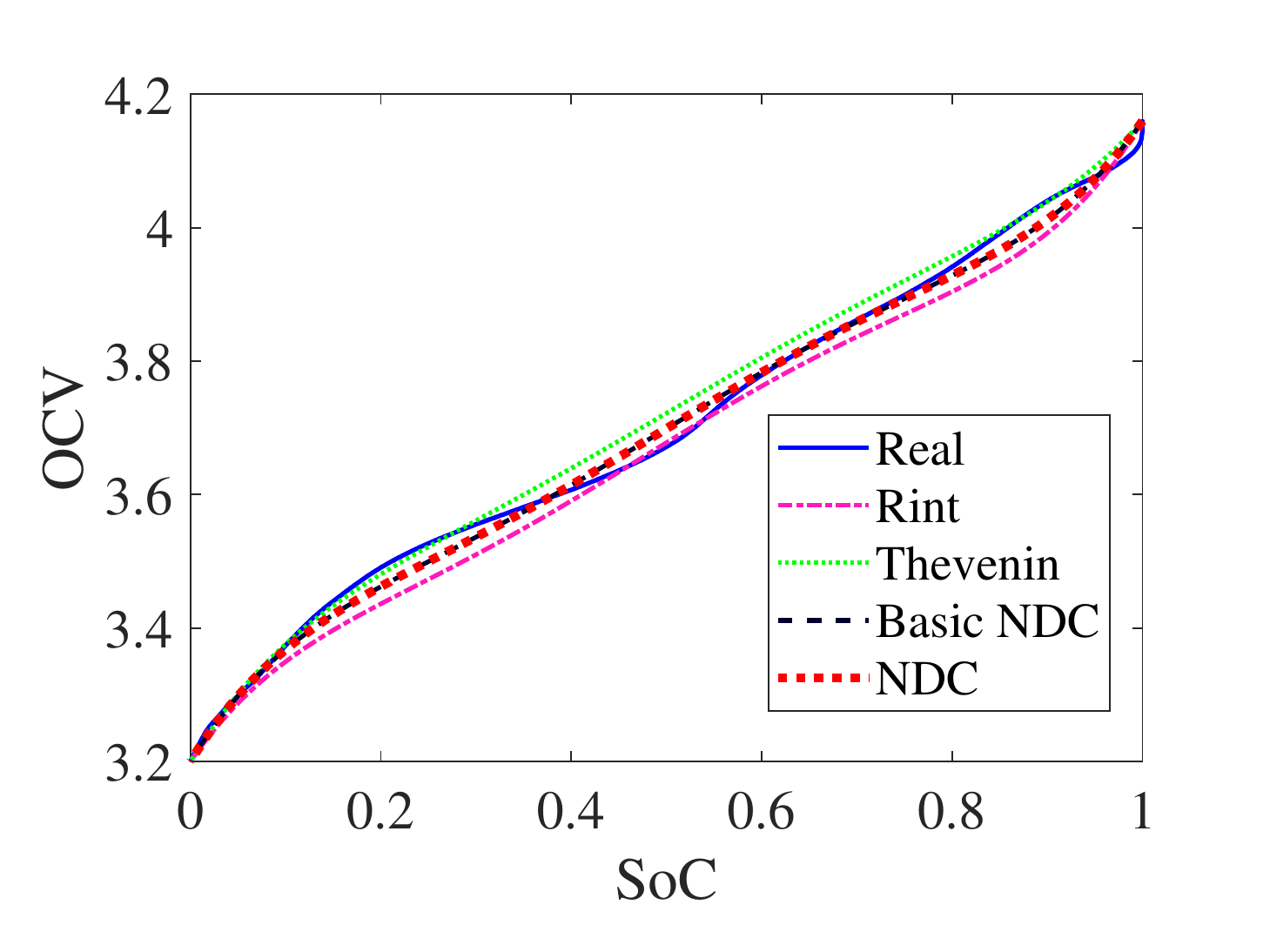}
\caption{Identification 2.0: identification of the SoC-OCV relation based on different models, compared to the truth.}
\label{Fig:SoC-OCV-Validation-Multimodel}
\vspace{0mm}
\end{figure}

Another evaluation of interest is about the SoC-OCV relation. As mentioned earlier, the 2.0   approach can estimate all the parameters, including the function $h(\cdot)$. This allows one to write the SoC-OCV function directly based on the identified $h(\cdot)$ as it   also characterizes the SoC-OCV relation. That is,
\begin{align}\nonumber
{\rm OCV} & = 3.2+2.32\cdot{\rm SoC}-8.15\cdot{\rm SoC}^2+19.345\cdot{\rm SoC}^3\\ \nonumber
& \quad-20.78\cdot{\rm SoC}^4+8.222\cdot{\rm SoC}^5.
\end{align}
Identification of the other three models can also lead to estimation of this function. 
Figure~\ref{Fig:SoC-OCV-Validation-Multimodel} compares them with the benchmark shown in Figure~\ref{Fig:SoC-OCV-Validation}, which is obtained experimentally by discharging the cell using a small current of 0.1 A. It is obvious that the SoC-OCV curves obtained in the identification of the NDC model and its basic version are closer to the benchmark overall. This  further shows the benefit of the NDC model as well as the efficacy of the 2.0   approach. 

\vspace{3mm}

Summing up the above validation results, one can draw the following observations:
\begin{itemize}

\item The NDC model is the most competent among the four considered models for grasping and predicting a battery's dynamic behavior, justifying its validity and soundness.

\item The basic NDC model can   offer  fitting and prediction accuracy almost comparable to that of the full model. It thus can be well qualified if a practitioner wants to use a simpler NDC model yet without  much loss of accuracy.

\item The 2.0 identification approach is effective in estimating all the parameters of the NDC model as well as the Rint and Thevenin models in one shot from variable-current-based data profiles. It can not only ease the cost of identification considerably but also  provide  on-demand model availability potentially in practice.

\end{itemize}



\section{Conclusion}\label{Sec:Conclusion}

The growing   importance of  real-time battery management has imposed a pressing demand for battery models with high fidelity and low complexity, making ECMs a popular choice in this field. The double-capacitor model is emerging as a favorable ECM for diverse applications,    promising several   advantages in capturing a battery's dynamics. However, its  linear structure intrinsically  hinders a characterization of a battery's nonlinear phenomena. To thoroughly improve this model, this paper   proposed to modify its original structure by adding  a nonlinear-mapping-based voltage source and a serial RC circuit. This development was justified through an analogous comparison with the SPM. Furthermore, two offline parameter estimation approaches, which were named 1.0 and  2.0, respectively,  were designed to identify the model from current/voltage data. The 1.0 approach considers the constant-current charging/discharging scenarios, determining the SoC-OCV relation first and then estimating the impedance and capacitance parameters. With the observation that the NDC model has a Wiener-type structure, the 2.0 approach was derived from the  Wiener perspective. As the first of its kind, it leverages the notion of  MAP  to address the issue of local minima that may reduce or damage the performance of the nonlinear Wiener system identification. It well lends itself to the variable-current charging/discharging scenarios and can desirably estimate all the parameters in one shot.  
The experimental evaluation  demonstrated that the NDC model outperformed the popularly used  Rint and Thevenin models  in predicting a battery's behavior, in addition to showing  the effectiveness of the identification approaches for  extracting parameters. Our future work will include: 1) enhancing the NDC model further to account for the effects of temperature and include  the voltage hysteresis,  2) investigating optimal input design for the model, and 3) building new battery estimation and control designs based on the  model.

\balance
\bibliographystyle{IEEEtran}

\bibliography{bibliography} 

\begin{thebibliography}{10}
\providecommand{\url}[1]{#1}
\csname url@samestyle\endcsname
\providecommand{\newblock}{\relax}
\providecommand{\bibinfo}[2]{#2}
\providecommand{\BIBentrySTDinterwordspacing}{\spaceskip=0pt\relax}
\providecommand{\BIBentryALTinterwordstretchfactor}{4}
\providecommand{\BIBentryALTinterwordspacing}{\spaceskip=\fontdimen2\font plus
\BIBentryALTinterwordstretchfactor\fontdimen3\font minus
  \fontdimen4\font\relax}
\providecommand{\BIBforeignlanguage}[2]{{%
\expandafter\ifx\csname l@#1\endcsname\relax
\typeout{** WARNING: IEEEtran.bst: No hyphenation pattern has been}%
\typeout{** loaded for the language `#1'. Using the pattern for}%
\typeout{** the default language instead.}%
\else
\language=\csname l@#1\endcsname
\fi
#2}}
\providecommand{\BIBdecl}{\relax}
\BIBdecl

\bibitem{Johnson:NREL:2000}
V.~H. Johnson and A.~A. Pesaran, ``Temperature-dependent battery models for
  high-power lithium-lon batteries,'' National Renewable Energy Laboratory,
  Tech. Rep. NREL/CP-540-28716, 2000.

\bibitem{DFN:JTES:1993}
M.~Doyle, T.~F. Fuller, and J.~Newman, ``Modeling of galvanostatic charge and
  discharge of the lithium/polymer/insertion cell,'' \emph{Journal of The
  Electrochemical Society}, vol. 140, no.~6, pp. 1526--1533, 1993.

\bibitem{Forman:JPS:2012}
J.~C. Forman, S.~J. Moura, J.~L. Stein, and H.~K. Fathy, ``Genetic
  identification and fisher identifiability analysis of the
  {Doyle–Fuller–Newman} model from experimental cycling of a
  {L}i{F}e{PO}$_4$ cell,'' \emph{Journal of Power Sources}, vol. 210, pp. 263
  -- 275, 2012.

\bibitem{Chaturvedi:CSM:2010}
N.~Chaturvedi, R.~Klein, J.~Christensen, J.~Ahmed, and A.~Kojic, ``Algorithms
  for advanced battery-management systems,'' \emph{IEEE Control Systems
  Magazine}, vol.~30, no.~3, pp. 49--68, 2010.

\bibitem{Guo:JES:2011}
M.~Guo, G.~Sikha, and R.~E. White, ``Single-particle model for a lithium-ion
  cell: Thermal behavior,'' \emph{Journal of The Electrochemical Society}, vol.
  158, no.~2, pp. A122--A132, 2011.

\bibitem{Northrop:JES:2014}
P.~W.~C. Northrop, B.~Suthar, V.~Ramadesigan, S.~Santhanagopalan, R.~D. Braatz,
  and V.~R. Subramanian, ``Efficient simulation and reformulation of
  lithium-ion battery models for enabling electric transportation,''
  \emph{Journal of The Electrochemical Society}, vol. 161, no.~8, pp.
  E3149--E3157, 2014.

\bibitem{Zou:TCST:2016}
C.~{Zou}, C.~{Manzie}, and D.~{Ne\v{s}i\'{c}}, ``A framework for simplification
  of {PDE}-based lithium-ion battery models,'' \emph{IEEE Transactions on
  Control Systems Technology}, vol.~24, no.~5, pp. 1594--1609, 2016.

\bibitem{Hu:Mechatronics:2018}
X.~{Hu}, D.~{Cao}, and B.~{Egardt}, ``Condition monitoring in advanced battery
  management systems: Moving horizon estimation using a reduced electrochemical
  model,'' \emph{IEEE/ASME Transactions on Mechatronics}, vol.~23, no.~1, pp.
  167--178, 2018.

\bibitem{Randles:DFS:1947}
J.~E.~B. Randles, ``Kinetics of rapid electrode reactions,'' \emph{Discussions
  of the Faraday Society}, vol.~1, pp. 11--19, Mar. 1947.

\bibitem{Zia:Springer:2016}
A.~I. Zia and S.~C. Mukhopadhyay, \emph{Impedance Spectroscopy and Experimental
  Setup}.\hskip 1em plus 0.5em minus 0.4em\relax Cham, Switzerland: Springer
  International Publishing, 2016, pp. 21--37.

\bibitem{Mousavi:RSER:2014}
S.~M. G. and M.~Nikdel, ``Various battery models for various simulation studies
  and applications,'' \emph{Renewable and Sustainable Energy Reviews}, vol.~32,
  pp. 477--485, 2014.

\bibitem{He:Energies:2011}
H.~He, R.~Xiong, and J.~Fan, ``Evaluation of lithium-ion battery equivalent
  circuit models for state of charge estimation by an experimental approach,''
  \emph{Energies}, vol.~4, pp. 582--598, 2011.

\bibitem{Plett:Artech:2015}
G.~L. Plett, \emph{Battery Management Systems, Volume 1: Battery
  Modeling}.\hskip 1em plus 0.5em minus 0.4em\relax Artech House, 2015.

\bibitem{Weng:JPS:2014}
C.~Weng, J.~Sun, and H.~Peng, ``A unified open-circuit-voltage model of
  lithium-ion batteries for state-of-charge estimation and state-of-health
  monitoring,'' \emph{Journal of Power Sources}, vol. 258, pp. 228--237, 2014.

\bibitem{Lin:JPS:2014}
X.~Lin, H.~E. Perez, S.~Mohan, J.~B. Siegel, A.~G. Stefanopoulou, Y.~Ding, and
  M.~P. Castanier, ``A lumped-parameter electro-thermal model for cylindrical
  batteries,'' \emph{Journal of Power Sources}, vol. 257, pp. 1--11, 2014.

\bibitem{gholizadeh2014estimation}
M.~Gholizadeh and F.~R. Salmasi, ``Estimation of state of charge, unknown
  nonlinearities, and state of health of a lithium-ion battery based on a
  comprehensive unobservable model,'' \emph{IEEE Transactions on Industrial
  Electronics}, vol.~61, no.~3, pp. 1335--1344, 2014.

\bibitem{Perez:TVT:2017}
H.~E. Perez, X.~Hu, S.~Dey, and S.~J. Moura, ``Optimal charging of {L}i-ion
  batteries with coupled electro-thermal-aging dynamics,'' \emph{IEEE
  Transactions on Vehicular Technology}, vol.~66, no.~9, pp. 7761--7770, 2017.

\bibitem{Fang:CEP:2014}
H.~Fang, Y.~Wang, Z.~Sahinoglu, T.~Wada, and S.~Hara, ``State of charge
  estimation for lithium-ion batteries: An adaptive approach,'' \emph{Control
  Engineering Practice}, vol.~25, pp. 45--54, 2014.

\bibitem{Plett:JPS:2004}
G.~L. Plett, ``Extended {K}alman filtering for battery management systems of
  {L}i{PB}-based {HEV} battery packs: Part 2. {M}odeling and identification,''
  \emph{Journal of Power Sources}, vol. 134, no.~2, pp. 262--276, 2004.

\bibitem{Hu:JPS:2012}
X.~Hu, S.~Li, and H.~Peng, ``A comparative study of equivalent circuit models
  for {L}i-ion batteries,'' \emph{Journal of Power Sources}, vol. 198, pp.
  359--367, 2012.

\bibitem{li2018practical}
K.~Li, F.~Wei, K.~J. Tseng, and B.-H. Soong, ``A practical lithium-ion battery
  model for state of energy and voltage responses prediction incorporating
  temperature and ageing effects,'' \emph{IEEE Transactions on Industrial
  Electronics}, vol.~65, no.~8, pp. 6696--6708, 2018.

\bibitem{lee2018temperature}
K.-T. Lee, M.-J. Dai, and C.-C. Chuang, ``Temperature-compensated model for
  lithium-ion polymer batteries with extended {K}alman filter state-of-charge
  estimation for an implantable charger,'' \emph{IEEE Transactions on
  Industrial Electronics}, vol.~65, no.~1, pp. 589--596, 2018.

\bibitem{Chen:TEC:2006}
M.~Chen and G.~A. Rincon-Mora, ``Accurate electrical battery model capable of
  predicting runtime and {I}-{V} performance,'' \emph{IEEE Transactions on
  Energy Conversion}, vol.~21, no.~2, pp. 504--511, 2006.

\bibitem{Kim:TEC:2011}
T.~Kim and W.~Qiao, ``A hybrid battery model capable of capturing dynamic
  circuit characteristics and nonlinear capacity effects,'' \emph{IEEE
  Transactions on Energy Conversion}, vol.~26, no.~4, pp. 1172--1180, 2011.

\bibitem{Johnson:JPS:2002}
V.~Johnson, ``Battery performance models in {ADVISOR},'' \emph{Journal of Power
  Sources}, vol. 110, no.~2, pp. 321--329, 2002.

\bibitem{Fang:TCST:2017}
H.~Fang, Y.~Wang, and J.~Chen, ``Health-aware and user-involved battery
  charging management for electric vehicles: Linear quadratic strategies,''
  \emph{IEEE Transactions on Control Systems Technology}, vol.~25, no.~3, pp.
  911--923, 2017.

\bibitem{Fang:JES:2018}
H.~Fang, C.~Depcik, and V.~Lvovich, ``Optimal pulse-modulated lithium-ion
  battery charging: Algorithms and simulation,'' \emph{Journal of Energy
  Storage}, vol.~15, pp. 359--367, 2018.

\bibitem{schweighofer2003modeling}
B.~Schweighofer, K.~M. Raab, and G.~Brasseur, ``Modeling of high power
  automotive batteries by the use of an automated test system,'' \emph{IEEE
  Transactions on Instrumentation and Measurement}, vol.~52, no.~4, pp.
  1087--1091, 2003.

\bibitem{abu2004rapid}
S.~Abu-Sharkh and D.~Doerffel, ``Rapid test and non-linear model
  characterisation of solid-state lithium-ion batteries,'' \emph{Journal of
  Power Sources}, vol. 130, no.~1, pp. 266--274, 2004.

\bibitem{dubarry2007development}
M.~Dubarry and B.~Y. Liaw, ``Development of a universal modeling tool for
  rechargeable lithium batteries,'' \emph{Journal of Power Sources}, vol. 174,
  no.~2, pp. 856--860, 2007.

\bibitem{he2011state}
H.~He, R.~Xiong, X.~Zhang, F.~Sun, and J.~Fan, ``State-of-charge estimation of
  the lithium-ion battery using an adaptive extended {K}alman filter based on
  an improved {T}hevenin model,'' \emph{IEEE Transactions on Vehicular
  Technology}, vol.~60, no.~4, pp. 1461--1469, 2011.

\bibitem{tian2014modified}
Y.~Tian, B.~Xia, W.~Sun, Z.~Xu, and W.~Zheng, ``A modified model based state of
  charge estimation of power lithium-ion batteries using unscented {K}alman
  filter,'' \emph{Journal of Power Sources}, vol. 270, pp. 619--626, 2014.

\bibitem{mauracher1997dynamic}
P.~Mauracher and E.~Karden, ``Dynamic modelling of lead/acid batteries using
  impedance spectroscopy for parameter identification,'' \emph{Journal of Power
  Sources}, vol.~67, no. 1-2, pp. 69--84, 1997.

\bibitem{goebel2008prognostics}
K.~Goebel, B.~Saha, A.~Saxena, J.~R. Celaya, and J.~P. Christophersen,
  ``Prognostics in battery health management,'' \emph{IEEE Instrumentation \&
  Measurement Magazine}, vol.~11, no.~4, 2008.

\bibitem{Birkl:HEVC:2013}
C.~Birkl and D.~Howey, ``Model identification and parameter estimation for
  {L}i{F}e{PO}$_4$ batteries,'' in \emph{Proceedings of {IET} Hybrid and
  Electric Vehicles Conference}, 2013, pp. 1--6.

\bibitem{Hu:JPS:2013}
T.~Hu and H.~Jung, ``Simple algorithms for determining parameters of circuit
  models for charging/discharging batteries,'' \emph{Journal of Power Sources},
  vol. 233, pp. 14--22, 2013.

\bibitem{he2016parameter}
Z.~He, G.~Yang, and L.~Lu, ``A parameter identification method for dynamics of
  lithium iron phosphate batteries based on step-change current curves and
  constant current curves,'' \emph{Energies}, vol.~9, no.~6, p. 444, 2016.

\bibitem{yang2016improved}
J.~Yang, B.~Xia, Y.~Shang, W.~Huang, and C.~Mi, ``Improved battery parameter
  estimation method considering operating scenarios for {HEV/EV}
  applications,'' \emph{Energies}, vol.~10, no.~1, p.~5, 2016.

\bibitem{tian2017parameter}
N.~Tian, Y.~Wang, J.~Chen, and H.~Fang, ``On parameter identification of an
  equivalent circuit model for lithium-ion batteries,'' in \emph{Control
  Technology and Applications (CCTA), 2017 IEEE Conference on}.\hskip 1em plus
  0.5em minus 0.4em\relax IEEE, 2017, pp. 187--192.

\bibitem{feng2015online}
T.~Feng, L.~Yang, X.~Zhao, H.~Zhang, and J.~Qiang, ``Online identification of
  lithium-ion battery parameters based on an improved equivalent-circuit model
  and its implementation on battery state-of-power prediction,'' \emph{Journal
  of Power Sources}, vol. 281, pp. 192--203, 2015.

\bibitem{prasad2013model}
G.~K. Prasad and C.~D. Rahn, ``Model based identification of aging parameters
  in lithium ion batteries,'' \emph{Journal of Power Sources}, vol. 232, pp.
  79--85, 2013.

\bibitem{sitterly2011enhanced}
M.~Sitterly, L.~Y. Wang, G.~G. Yin, and C.~Wang, ``Enhanced identification of
  battery models for real-time battery management,'' \emph{IEEE Transactions on
  Sustainable Energy}, vol.~2, no.~3, pp. 300--308, 2011.

\bibitem{zhang2014multi}
L.~Zhang, L.~Wang, G.~Hinds, C.~Lyu, J.~Zheng, and J.~Li, ``Multi-objective
  optimization of lithium-ion battery model using genetic algorithm approach,''
  \emph{Journal of Power Sources}, vol. 270, pp. 367--378, 2014.

\bibitem{rahman2016electrochemical}
M.~A. Rahman, S.~Anwar, and A.~Izadian, ``Electrochemical model parameter
  identification of a lithium-ion battery using particle swarm optimization
  method,'' \emph{Journal of Power Sources}, vol. 307, pp. 86--97, 2016.

\bibitem{yu2017model}
Z.~Yu, L.~Xiao, H.~Li, X.~Zhu, and R.~Huai, ``Model parameter identification
  for lithium batteries using the coevolutionary particle swarm optimization
  method,'' \emph{IEEE Trans. Ind. Electron}, vol.~64, no.~7, pp. 5690--5700,
  2017.

\bibitem{Limoge:TCST:2018}
D.~W. {Limoge} and A.~M. {Annaswamy}, ``An adaptive observer design for
  real-time parameter estimation in lithium-ion batteries,'' \emph{IEEE
  Transactions on Control Systems Technology}, 2018, in press.

\bibitem{Hu:JPS:2011}
Y.~Hu and S.~Yurkovich, ``Linear parameter varying battery model identification
  using subspace methods,'' \emph{Journal of Power Sources}, vol. 196, no.~5,
  pp. 2913--2923, 2011.

\bibitem{Li:IJER:2014}
Y.~Li, C.~Liao, L.~Wang, L.~Wang, and D.~Xu, ``Subspace-based modeling and
  parameter identification of lithium-ion batteries,'' \emph{International
  Journal of Energy Research}, vol.~38, no.~8, pp. 1024--1038, 2014.

\bibitem{Relan:TCST:2017}
R.~{Relan}, Y.~{Firouz}, J.~{Timmermans}, and J.~{Schoukens}, ``Data-driven
  nonlinear identification of li-ion battery based on a frequency domain
  nonparametric analysis,'' \emph{IEEE Transactions on Control Systems
  Technology}, vol.~25, no.~5, pp. 1825--1832, 2017.

\bibitem{Rothenberger:JES:2015}
M.~J. Rothenberger, D.~J. Docimo, M.~Ghanaatpishe, and H.~K. Fathy, ``Genetic
  optimization and experimental validation of a test cycle that maximizes
  parameter identifiability for a {Li-ion} equivalent-circuit battery model,''
  \emph{Journal of Energy Storage}, vol.~4, pp. 156--166, 2015.

\bibitem{Park:ACC:2018}
S.~Park, D.~Kato, Z.~Gima, R.~Klein, and S.~Moura, ``Optimal input design for
  parameter identification in an electrochemical {Li-ion} battery model,'' in
  \emph{Proceedings of American Control Conference}, 2018, pp. 2300--2305.

\bibitem{giri2010block}
F.~Giri and E.-W. Bai, \emph{Block-{O}riented {N}onlinear {S}ystem
  {I}dentification}.\hskip 1em plus 0.5em minus 0.4em\relax London, UK:
  Springer, 2010, vol.~1.

\bibitem{Hagenblad:AUTO:2008}
A.~Hagenblad, L.~Ljung, and A.~Wills, ``Maximum likelihood identification of
  {Wiener} models,'' \emph{Automatica}, vol.~44, no.~11, pp. 2697 -- 2705,
  2008.

\bibitem{Vanbeylen:TSP:2009}
L.~Vanbeylen, R.~Pintelon, and J.~Schoukens, ``Blind maximum-likelihood
  identification of {W}iener systems,'' \emph{IEEE Transactions on Signal
  Processing}, vol.~57, no.~8, pp. 3017--3029, 2009.

\bibitem{Dey:TCST:2019}
S.~{Dey}, S.~{Mohon}, B.~{Ayalew}, H.~{Arunachalam}, and S.~{Onori}, ``A novel
  model-based estimation scheme for battery-double-layer capacitor hybrid
  energy storage systems,'' \emph{IEEE Transactions on Control Systems
  Technology}, vol.~27, no.~2, pp. 689--702, 2019.

\bibitem{mamun2018collective}
A.~Mamun, A.~Sivasubramaniam, and H.~Fathy, ``Collective learning of
  lithium-ion aging model parameters for battery health-conscious demand
  response in datacenters,'' \emph{Energy}, vol. 154, pp. 80--95, 2018.

\bibitem{andre2011characterization}
D.~Andre, M.~Meiler, K.~Steiner, C.~Wimmer, T.~Soczka-Guth, and D.~Sauer,
  ``Characterization of high-power lithium-ion batteries by electrochemical
  impedance spectroscopy. {I}. {E}xperimental investigation,'' \emph{Journal of
  Power Sources}, vol. 196, no.~12, pp. 5334--5341, 2011.

\bibitem{dubarry2009single}
M.~Dubarry, N.~Vuillaume, and B.~Y. Liaw, ``From single cell model to battery
  pack simulation for {L}i-ion batteries,'' \emph{Journal of Power Sources},
  vol. 186, no.~2, pp. 500--507, 2009.

\bibitem{Hagenblad:Thesis:1999}
A.~Hagenblad, ``Aspects of the identification of {Wiener} models,'' Ph.D.
  dissertation, Department of Electrical Engineering, Link\"{o}ping University,
  Sweden, 1999.

\bibitem{wright1999numerical}
S.~Wright and J.~Nocedal, \emph{Numerical {O}ptimization}.\hskip 1em plus 0.5em
  minus 0.4em\relax New York, US: Springer-Verlag New York, 1999, vol.~35, no.
  67--68.

\bibitem{VanDoren:IFAC:2009}
J.~F. {Van Doren}, S.~G. Douma, P.~M. {Van den Hof}, J.~D. Jansen, and O.~H.
  Bosgra, ``Identifiability: from qualitative analysis to model structure
  approximation,'' \emph{IFAC Proceedings Volumes of the 15th IFAC Symposium on
  System Identification}, vol.~42, no.~10, pp. 664 -- 669, 2009.

\bibitem{UDDS}
``The {EPA} {U}rban {D}ynamometer {D}riving {S}chedule ({UDDS}) [{O}nline],''
  Available:
  \url{https://www.epa.gov/sites/production/files/2015-10/uddscol.txt}.

\end{thebibliography}

%

\begin{IEEEbiography}
[{\includegraphics[width=1in,height=1.25in,clip,keepaspectratio]{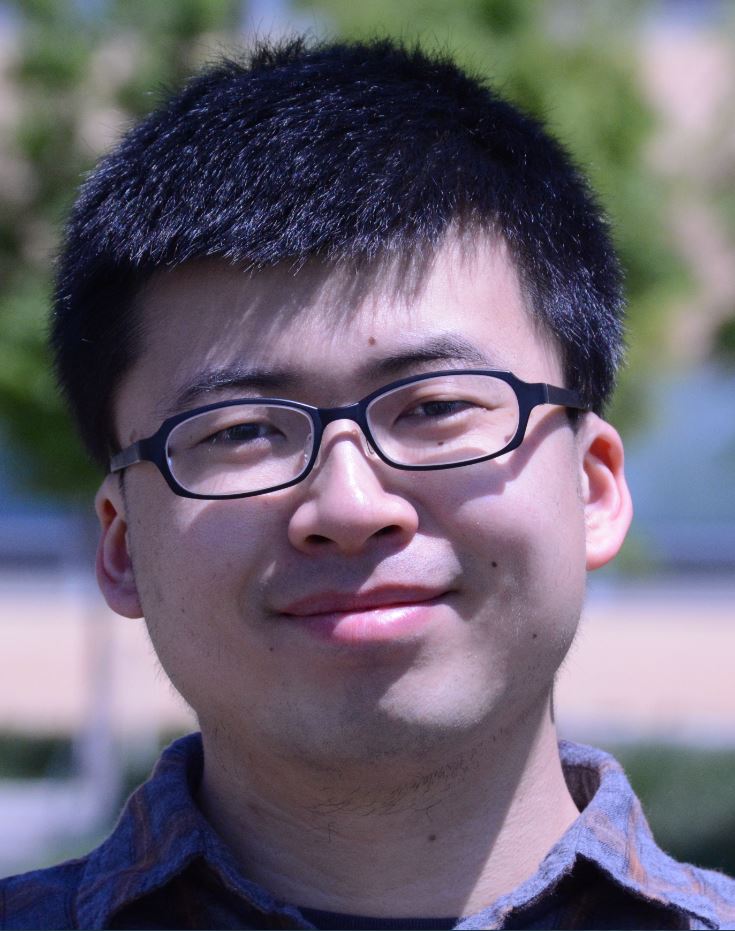}}]{Ning Tian}
received the B.Eng. and M.Sc. degrees in thermal engineering from Northwestern Polytechnic University, Xi’an, China, in 2012 and 2015.
Currently, he is a Ph.D. candidate at the University of Kansas, Lawrence, KS, USA. His research interests include control theory and its
application to advanced battery management.
\end{IEEEbiography}

\vspace{-0.3cm}

\begin{IEEEbiography}
[{\includegraphics[width=1in,height=1.25in,clip,keepaspectratio]{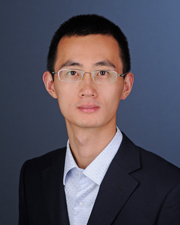}}]{Huazhen Fang}
(M’14) received the B.Eng. degree in computer science and technology from Northwestern Polytechnic University, Xi’an, China, in 2006, the M.Sc. degree in mechanical engineering from the University of Saskatchewan, Saskatoon, SK, Canada, in 2009, and the Ph.D. degree in mechanical engineering from the University of California, San Diego, CA, USA, in 2014.

He is currently an Assistant Professor of Mechanical Engineering with the University of Kansas, Lawrence, KS, USA. His research interests include control and estimation theory with application to energy management, cooperative robotics, and system prognostics.

Dr. Fang received the 2019 National Science Foundation CAREER Award and the awards of Outstanding Reviewer or Reviewers of the Year from {\em Automatica}, {\em IEEE Transactions on Cybernetics}, and {\em ASME Journal of Dynamic Systems, Measurement and Control}.
\end{IEEEbiography}

\begin{IEEEbiography}
[{\includegraphics[width=1in,height=1.25in,clip,keepaspectratio]{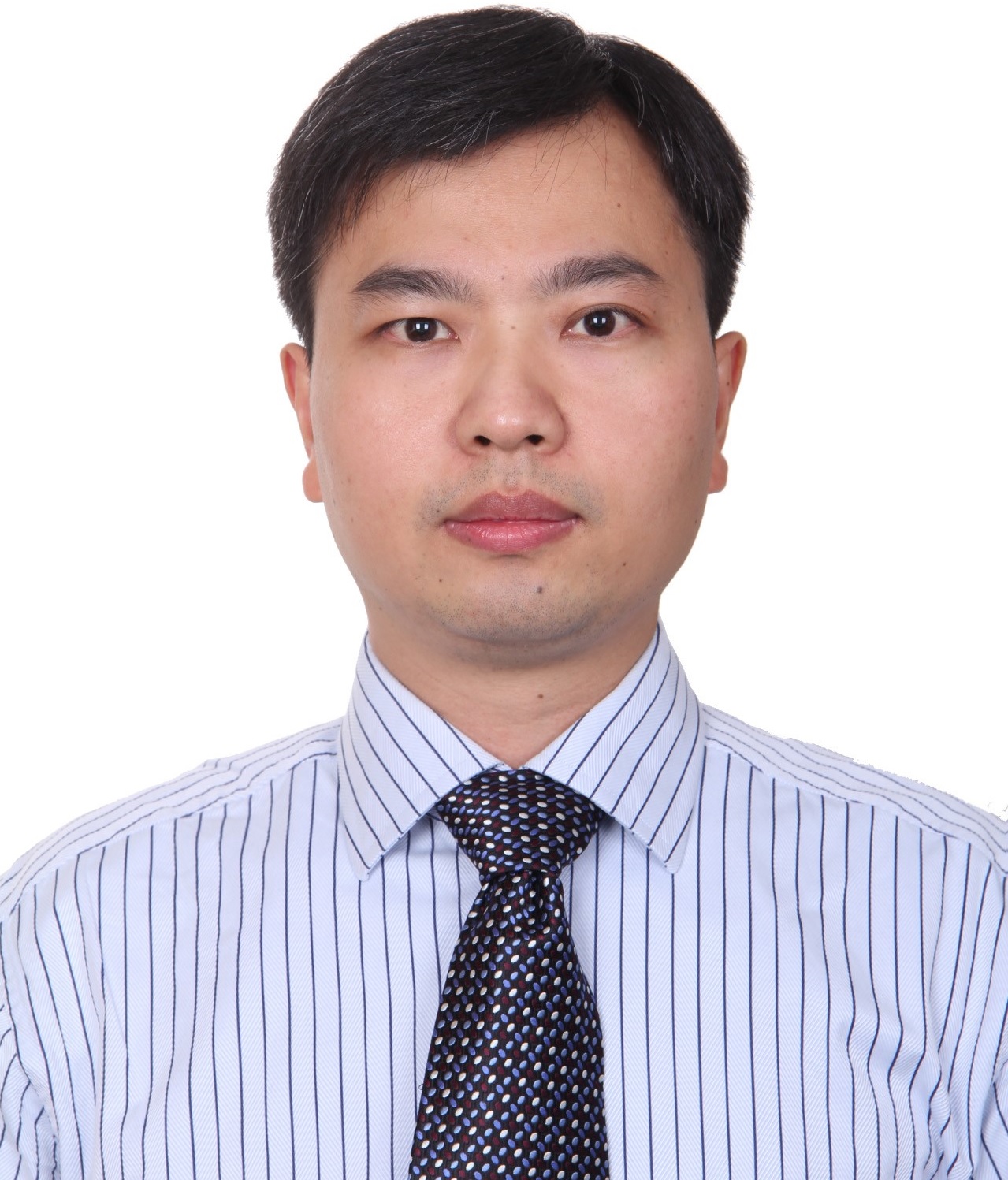}}]{Jian Chen}
(M’06-SM’10) received the B.E. degree in measurement and control technology and instruments and the M.E. degree in control science and engineering from Zhejiang University, Hangzhou, China, in 1998 and 2001, respectively, and the Ph.D. degree in electrical engineering from Clemson University, Clemson, SC, USA, in 2005. He was a Research Fellow with the University of Michigan, Ann Arbor, MI, USA, from 2006 to 2008, where he was involved in fuel cell modeling and control. In 2013, he joined the Department of Control Science and Engineering, Zhejiang University, where he is currently a Professor with the College of Control Science and Engineering. His research interests include modeling and control of fuel cell systems, vehicle control and intelligence, machine vision, and nonlinear control.
\end{IEEEbiography}

\vspace{-0.6cm}

\begin{IEEEbiography}
[{\includegraphics[width=1in,height=1.25in,clip,keepaspectratio]{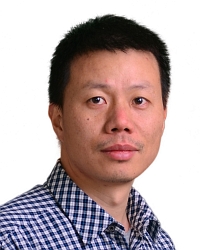}}]{Yebin Wang}
(M’10-SM’16) received the B.Eng. degree in Mechatronics Engineering from Zhejiang University, Hangzhou, China, in 1997, M.Eng. degree in Control Theory \& Control Engineering from Tsinghua University, Beijing, China, in 2001, and Ph.D. in Electrical Engineering from the University of Alberta, Edmonton, Canada, in 2008.

Dr. Wang has been with Mitsubishi Electric Research Laboratories in Cambridge, MA, USA, since 2009, and now is a Senior Principal Research Scientist. From 2001 to 2003 he was a Software Engineer, Project Manager, and Manager of R\&D Dept. in industries, Beijing, China. His research interests include nonlinear control and estimation, optimal control, adaptive systems and their applications including mechatronic systems.
\end{IEEEbiography}

\end{document}